%% file: ms.tex
\documentclass[12pt]{article}
\usepackage[margin=1in]{geometry}
\usepackage{setspace}
\usepackage{amsmath,amssymb,amsthm,mathtools}
\usepackage[numbers,sort&compress]{natbib}
\usepackage{booktabs,multirow,threeparttable,array}
\usepackage{graphicx,caption,subcaption}
\usepackage{enumitem}
\usepackage{algorithm,algpseudocode}
\usepackage{xcolor}
\usepackage{placeins}
\usepackage[colorlinks=true,linkcolor=blue!50!black,citecolor=blue!50!black,urlcolor=blue!50!black,hypertexnames=false]{hyperref}

\input{macros}

\newtheorem{theorem}{Theorem}
\newtheorem{lemma}{Lemma}
\newtheorem{corollary}{Corollary}
\newtheorem{proposition}{Proposition}
\newtheorem{assumption}{Assumption}
\newtheorem{condition}{Condition}
\theoremstyle{definition}

\newtheorem{remark}{Remark}

\newcommand{\anon}{0}

\begin{document}
\onehalfspacing

\begin{center}
{\Large\bf Efficient inference for fairness-constrained\\[2pt] sequential decision policies}

\vspace{0.5cm}
\ifnum\anon=1
(Author names and affiliations omitted for blind review)
\else
Yi-Hui Zhou\\[2pt]
{\it Department of Statistics and Bioinformatics Research Center,\\
North Carolina State University, Raleigh, NC 27695, USA}\\[2pt]
E-mail: \texttt{yihui\_zhou@ncsu.edu}
\fi
\end{center}

\begin{abstract}
\noindent
Adaptive decision systems in education and health choose interventions sequentially and are often
required not to leave a protected group too far behind. We study the fairness-constrained optimal
value: the largest population value attainable by a policy in a smooth parametric class subject to
constraints that are smooth functions of group-specific counterfactual functionals, such as value
gaps, floors, max--min welfare and budgets. For multi-stage, non-Markov data under sequential
ignorability, a sensitivity analysis of the Karush--Kuhn--Tucker system shows that this value is
pathwise differentiable, with efficient influence function equal to the Lagrangian combination of
the influence functions of the group values. The same analysis gives the influence functions of the
optimal policy, the multiplier and the group values it delivers. A cross-fitted doubly robust
estimator is efficient despite its data-dependent feasible set and yields confidence intervals for
the constrained value and the price of fairness. When the constraint binds, any plug-in fair policy
violates the population constraint, by a margin that vanishes as the sample grows, with probability
tending to one half; tightening the estimated constraint by a one-sided confidence margin certifies
fairness at a small cost in value. Exact checks, simulations and a synthetic case study illustrate
the theory.

\vspace{6pt}
\noindent{\it Keywords:} constrained optimization, dynamic treatment regimes, efficient influence
function, fairness, off-policy evaluation, price of fairness

\vspace{3pt}
\noindent{\it MSC 2020:} 62D20, 62G05, 62G20, 90C31
\end{abstract}

\section{Introduction}\label{sec:intro}

Intelligent tutoring platforms, mobile-health applications and adaptive social programmes choose
interventions sequentially on the basis of each individual's evolving history. Policies learned
from logged data can raise average outcomes, but they can also widen existing gaps between
demographic groups, for instance when a pedagogical action is most beneficial for students who
already perform well \citep{mitchell2021algorithmic,liu2018delayed,damour2020fairness}. A
decision maker who is required to protect a disadvantaged group therefore faces two statistical
questions: how much average performance can be attained by a policy that satisfies a given
fairness requirement, and how much is lost relative to the unconstrained optimum? Both quantities
are functionals of the unknown data-generating distribution, and both are defined through an
optimization problem whose feasible set is itself unknown and must be estimated from the same
data.

This paper develops efficiency theory and inference for these quantities in multi-stage, possibly
non-Markov, sequential decision problems. Let $V(\theta)$ be the population value of a policy
$\pi_\theta$ in a smooth parametric class and let $V_g(\theta)$ be its value in group $g$. Our
running example is the \emph{fairness-constrained optimal value}
\begin{equation}\label{eq:intro}
\Vstar(\delta)=\max_{\theta\in\Theta}\ V(\theta)\quad\text{subject to}\quad V_A(\theta)\ge V_B(\theta)-\delta,
\end{equation}
which requires that the protected group $A$ is not left more than $\delta$ behind group $B$,
together with the \emph{price of fairness} $\PoF(\delta)=\max_\theta V(\theta)-\Vstar(\delta)$, the
absolute version of the notion of \citet{bertsimas2011price}. Our theory, however, is developed for a general class of fairness
criteria (Section~\ref{sec:criteria}): any objective and constraints that are smooth functions of
group-specific counterfactual functionals. This class includes two-sided (absolute) value gaps,
ratio and floor constraints, ``no-harm'' constraints relative to the status quo, Rawlsian
max--min welfare, parity of intervention rates and resource constraints.

\paragraph{Related work.}
The efficient influence function (EIF) of the value of a fixed dynamic regime or stochastic policy
is classical \citep{robins1986new,murphy2001marginal,vanderlaan2003unified,bang2005doubly,zhang2013robust,tsiatis2019dynamic}
and underlies doubly robust off-policy evaluation in reinforcement learning
\citep{jiang2016doubly,thomas2016data,kallus2020double}, including non-Markov decision processes
\citep{kallus2020double}; it is the building block of our analysis (Lemma~\ref{lem:eif-fixed}).
Inference for the unconstrained optimal value was developed by \citet{vanderlaan2015targeted} and
\citet{luedtke2016statistical}, who also characterise the nonregularity that arises when optimal
rules are not unique; see also \citet{chakraborty2014inference} and \citet{laber2014dynamic}.
Optimal rules under constraints have been studied for resource and cost limits
\citep{luedtke2016resource,qiu2022individualized} and benefit--risk trade-offs
\citep{wang2018learning}; \citet{luedtke2016resource} derive the canonical gradient of the
resource-constrained optimal value for single-stage, unrestricted rules, which contains a multiplier
term. In stochastic programming, the first-order behaviour of an optimal value under expected-value
constraints is governed by the Lagrangian at the optimum \citep{shapiro1991asymptotic,shapiro2014lectures}.
Theorem~\ref{thm:eif}(i) extends this Lagrangian principle to group-level counterfactual functionals in
multi-stage problems. Beyond it, we derive the influence functions of the optimal policy, the
multiplier and the group values delivered by the fair policy, establish efficient cross-fitted
doubly robust estimation with an estimated feasible set, and characterise the feasibility of the
estimated fair policy.

Fairness in policy learning has been formalised through path-specific constraints on how decisions
depend on a sensitive attribute \citep{nabi2018fair,nabi2019learning}, equality-minded social welfare
functions \citep{kitagawa2021equality}, Pareto-optimal policies that minimise unfairness
\citep{viviano2024fair}, quantile-based criteria for individualised treatment rules
\citep{fang2023fairness}, fairness constraints on treatment-effect estimators used for targeting
\citep{kim2023fair}, action and value fairness in off-policy learning \citep{frauen2024fair}, access
parity in encouragement designs \citep{zhou2023encouragement}, and fairness notions for sequential and
reinforcement learning \citep{jabbari2017fairness,creager2020causal}. These works mostly study
learning rules with welfare or regret guarantees, often in single-stage problems; we focus on
efficient inference for multi-stage problems with group-level constraints.

\paragraph{Contributions.}
\begin{enumerate}[leftmargin=*,itemsep=2pt]
\item \emph{A general class of fairness criteria} (Section~\ref{sec:criteria}, Table~\ref{tab:criteria}):
objectives and constraints that are smooth functions of group-specific counterfactual functionals,
covering gaps, ratios, floors, no-harm constraints, max--min welfare, rate parity, budgets and
inequality-averse objectives.
\item \emph{Efficiency theory for the constrained optimum} (Theorem~\ref{thm:eif},
Corollary~\ref{cor:other}). Under standard second-order conditions, $\Vstar$ is pathwise differentiable,
and its EIF is the Lagrangian combination of the EIFs of the group functionals at the optimum; for
\eqref{eq:intro} it is $\varphi_{+,\thstar}+\lamstar(\varphi_{A,\thstar}-\varphi_{B,\thstar})$, where
$\varphi_{g,\thstar}$ is the EIF of the value of the optimal policy in group $g$ and $\lamstar$ is the
Lagrange multiplier. The sensitivity
analysis of the Karush--Kuhn--Tucker (KKT) system also gives the EIFs of the optimal policy $\thstar$,
the multiplier $\lamstar$, the group values delivered by the fair policy and the price of fairness.
Exact numerical differentiation confirms all formulas (Section~\ref{sec:exact}).
\item \emph{Efficient estimation with an estimated feasible set} (Theorem~\ref{thm:est}). The
cross-fitted doubly robust plug-in estimator computed by Algorithm~\ref{alg:main} is asymptotically
linear with influence function $\phistar$, hence regular and efficient, and yields Wald intervals for
$\Vstar(\delta)$ and, when the constraint binds, for $\PoF(\delta)$; under an additional condition the
same holds for $\lamstar$ and the group values at the optimum.
\item \emph{Feasibility and certification} (Proposition~\ref{prop:cert}). When the constraint binds,
a plug-in fair policy, which satisfies the estimated constraint exactly, violates the population
constraint with probability tending to $1/2$, by an amount of order $n^{-1/2}$. Tightening the
estimated constraint by a one-sided confidence margin yields feasibility with asymptotic probability
$1-\alpha$, at a value loss of order $\lamstar n^{-1/2}$.
\end{enumerate}

Section~\ref{sec:setup} introduces the setting and the fairness criteria;
Section~\ref{sec:ident} recalls identification and the fixed-policy EIF;
Section~\ref{sec:theory} contains the efficiency theory; Section~\ref{sec:estimation} the estimator,
algorithm and inference; Sections~\ref{sec:sim} and \ref{sec:case} report numerical studies; and
Section~\ref{sec:discussion} concludes. Proofs, details of the exact numerical verification and
additional results are in the Supplementary Material.

\section{Setting and fairness-constrained estimands}\label{sec:setup}

\subsection{Data and policies}

For each individual we observe $O=(H_1,A_1,Y_1,X_2,A_2,Y_2,\ldots,X_T,A_T,Y_T,X_{T+1})$ over a
fixed number $T$ of decision points. The baseline history $H_1=(G,X_1)$ contains a group label
$G\in\cG=\{1,\ldots,K\}$ and covariates $X_1$; the history before decision $t$ is
$H_t=(H_{t-1},A_{t-1},Y_{t-1},X_t)$; $A_t$ takes values in a finite set $\Aset$; and $Y_t$ is an
intermediate outcome. No Markov structure is imposed: $X_{t+1}$ and $Y_t$ may depend on the whole
history. Outcomes of interest are known functions $R^{(m)}=r_m(H_{T+1})$, $m=0,\ldots,M$, of the
complete history; $R\equiv R^{(0)}$ is the primary outcome (for example end-of-course mastery),
and auxiliary outcomes such as the fraction of decision points at which an intervention is given,
$R^{(1)}=T^{-1}\sum_t\1\{A_t=1\}$, are used to express constraints on the decisions themselves. We
observe $n$ independent copies $O_1,\ldots,O_n$ of $O\sim P_0$. Data are collected under a
\emph{logging policy} $\mu_t(a\mid h)=P_0(A_t=a\mid H_t=h)$, which may be known (adaptive platforms
usually record their randomisation probabilities) or unknown.

A stochastic policy $\pi=(\pi_1,\ldots,\pi_T)$ assigns to each history a distribution
$\pi_t(\cdot\mid H_t)$ on $\Aset$. We consider a parametric class $\Pi_\Theta=\{\pi_\theta:\theta\in\Theta\}$,
$\Theta\subset\R^d$ compact, whose members are smooth in $\theta$ (Condition~\ref{cond:smooth}
below). In the running example there are two groups, labelled $A$ (protected) and $B$. The policy may or may not use $G$. A leading example, used in
Sections~\ref{sec:sim}--\ref{sec:case}, is the \emph{soft-threshold} class
\begin{equation}\label{eq:softthreshold}
\pi_{\theta,t}(1\mid H_t)=\expit\{\kappa(S_t-\theta_{s(t)})\},
\end{equation}
where $S_t$ is a history summary (such as recent average mastery), $\theta_{s(t)}$ is a
stage-specific threshold and $\kappa>0$ is a fixed temperature. Randomised policies of this type
are common in adaptive systems because they preserve exploration and positivity for future data
collection \citep{luckett2020estimating,liao2021off}; smoothness in $\theta$ is also what makes the
constrained optimum a regular statistical functional (see Section~\ref{sec:discussion} for unrestricted
classes).

\subsection{Counterfactual group functionals}

Let $H^{\pi}_{T+1}$ denote the counterfactual history that would arise if actions were drawn
sequentially from $\pi$, with randomisation independent of everything else, and define for
$g\in\cG$ and the whole population (indexed by ``$+$'')
\[
V^{(m)}_g(\theta)=\E\{r_m(H^{\pi_\theta}_{T+1})\mid G=g\},\qquad
V^{(m)}_+(\theta)=\E\{r_m(H^{\pi_\theta}_{T+1})\}=\textstyle\sum_g p_gV_g^{(m)}(\theta),
\]
where $p_g=P_0(G=g)$. We write $V_g=V_g^{(0)}$, $V=V_+^{(0)}$, and $D_g=V_g^{(1)}$ and $D_+=V_+^{(1)}$
for the expected intervention rates in group $g$ and overall. We also use \emph{status-quo} functionals $s_g=\E(R\mid G=g)$, the
group values under the logging policy, and the proportions $p_g$. Collect all functionals that
enter a fairness criterion into a vector $\bV(\theta)=\bV(\theta;P_0)\in\R^q$ (its dependence on the
distribution $P$ is made precise in Section~\ref{sec:ident}).

\subsection{The fairness-constrained optimal value}\label{sec:criteria}

A fairness criterion is specified by known smooth functions $c_0,\ldots,c_J$ (see
Condition~\ref{cond:smooth}) of a decision variable $\vth=(\theta,u)\in\Theta\times\cU$ and of
$\bv\in\R^q$, whose components $v_+,v_g,d_g,\ldots$ correspond to $V,V_g,D_g,\ldots$; here $u\in\cU\subset\R^{d_u}$ is an optional auxiliary variable (used for max--min
welfare) with $\cU$ compact, and $d_u=0$ when it is not used. With $C_j(\vth;P)=c_j\{\vth,\bV(\theta;P)\}$, the estimand is
\begin{equation}\label{eq:general}
\Vstar(P)=\max_{\vth\in\Theta\times\cU}C_0(\vth;P)\quad\text{subject to}\quad C_j(\vth;P)\ge0,\ j=1,\ldots,J.
\end{equation}
The running example \eqref{eq:intro} has $c_0=v_+$ and $c_1=v_A-v_B+\delta$; the unconstrained
optimal value $V^\star_{\rm unc}=\max_\theta V(\theta)$ corresponds to $J=0$, and
$\PoF(\delta)=V^\star_{\rm unc}-\Vstar(\delta)\ge0$. For objectives other than the population value
we use the same name for the loss in population value, $V^\star_{\rm unc}-V(\thstar)$, whose influence
function follows from Corollary~\ref{cor:other}.

\begin{table}[t]
\centering
\caption{Fairness criteria covered by \eqref{eq:general}. In each case the efficient influence
function of $\Vstar$ follows from Theorem~\ref{thm:eif}; $\varphi$'s denote influence functions of
the group functionals at the optimum (Lemma~\ref{lem:eif-fixed}), $\varphi^D_g$ and $\varphi^D_+$ those of
$D_g$ and $D_+$, and $\varphi^s_g$ that of $s_g$. Multipliers of inactive constraints are zero. Row (d) with
$\delta=1/5$ is the four-fifths rule and requires $v_B>0$; row (k) requires $v_g>0$, and its influence function also involves
those of the proportions $p_g$.}
\label{tab:criteria}
\small
\begin{tabular}{@{}>{\raggedright\arraybackslash}p{2.6cm}>{\raggedright\arraybackslash}p{4.1cm}>{\raggedright\arraybackslash}p{4.6cm}>{\raggedright\arraybackslash}p{3.4cm}@{}}
\toprule
Criterion & Objective / constraints & EIF of $\Vstar$ & Related work\\
\midrule
(a) Unconstrained & $\max V$ & $\varphi_+$ & \citep{murphy2001marginal,luedtke2016statistical}\\
(b) One-sided gap & $v_A-v_B+\delta\ge0$ & $\varphi_++\lambda(\varphi_A-\varphi_B)$ & running example\\
(c) Absolute gap & $\delta\mp(v_A-v_B)\ge0$ & $\varphi_++(\lambda_2-\lambda_1)(\varphi_A-\varphi_B)$ & statistical parity of outcomes\\
(d) Ratio & $v_A-(1-\delta)v_B\ge0$ & $\varphi_++\lambda\{\varphi_A-(1-\delta)\varphi_B\}$ & ``four-fifths'' rule ($v_B>0$)\\
(e) $K$-group gaps & $\delta-(v_g-v_h)\ge0$, $g\neq h$ & $\varphi_++\sum_{g\ne h}\lambda_{gh}(\varphi_h-\varphi_g)$ & subgroup fairness; \citep{kearns2018preventing}\\
(f) Floors & $v_g-\nu_g\ge0$ & $\varphi_++\sum_g\lambda_g\varphi_g$ & minimum guarantees\\
(g) No harm & $v_g-s_g\ge0$ & $\varphi_++\sum_g\lambda_g(\varphi_g-\varphi^s_g)$ & status-quo protection\\
(h) Max--min & $\max u$ s.t.\ $v_g-u\ge0$ & $\sum_g\lambda_g\varphi_g$, $\sum_g\lambda_g=1$ & \citep{rawls1971theory,frauen2024fair}\\
(i) Rate parity & $\delta\mp(D_A-D_B)\ge0$ & $\varphi_++(\lambda_2-\lambda_1)(\varphi^D_A-\varphi^D_B)$ & decision parity; \citep{zhou2023encouragement}\\
(j) Budget & $b-D_+\ge0$ & $\varphi_+-\lambda\varphi^D_+$ & \citep{luedtke2016resource}\\
(k) Inequality aversion & $\max(\sum_gp_gv_g^{1-\gamma})^{1/(1-\gamma)}$ & chain rule (Theorem~\ref{thm:eif}) & \citep{atkinson1970measurement}\\
\bottomrule
\end{tabular}
\end{table}

Table~\ref{tab:criteria} lists criteria covered by \eqref{eq:general}. Three remarks place the
running example among them.

First, the one-sided gap (b) protects a group identified in advance as disadvantaged. The absolute
gap (c) consists of two smooth constraints, at most one of which is active when $\delta>0$; for
$\delta=0$ it is an equality constraint (Section~\ref{S-sec:remarks} of the Supplement). The choice
between (b) and (c) is normative rather than statistical, and the theory covers both.

Second, gap-type constraints (b)--(e) can be met by \emph{levelling down}, that is, by lowering the
value of the advantaged group. Floors, no-harm constraints and max--min welfare, (f)--(h), avoid this,
and they can be combined with gap constraints within \eqref{eq:general}; the case study in
Section~\ref{sec:case} compares these choices on the same data. Gap constraints are most natural when
one policy is applied to all groups, for instance because the protected attribute may not be used for
decisions, or when the groups share a resource such as an intervention budget. If each group could
receive its own policy and nothing were shared, the unconstrained optimum would already be optimal
within each group, and a gap constraint could be met only by levelling down. Rate parity (i)
constrains the decisions rather than the outcomes, in the spirit of statistical parity in
classification \citep{dwork2012fairness}.

Third, individual-level and path-specific criteria
\citep{kusner2017counterfactual,nabi2018fair,nabi2019learning} restrict how the policy may depend on
$G$; they are imposed through the policy class $\Pi_\Theta$ and can be combined with
\eqref{eq:general}. Other notions are outside the class. Error-rate criteria
\citep{hardt2016equality,chouldechova2017fair,mishler2021fairness} and principal fairness
\citep{imai2023principal} involve joint distributions of potential outcomes, which are not identified
under Assumption~\ref{ass:causal} alone. Rank-dependent welfare \citep{kitagawa2021equality} and
quantile criteria \citep{fang2023fairness} depend on the whole outcome distribution. For such
criteria the sensitivity argument of Section~\ref{sec:theory} still applies whenever the underlying
functionals are identified and pathwise differentiable, but their influence functions differ from
those in Lemma~\ref{lem:eif-fixed}.

\section{Identification and the fixed-policy influence function}\label{sec:ident}

\begin{assumption}\label{ass:causal}
(i) \emph{Consistency}: the observed history equals the counterfactual history under the observed
actions. (ii) \emph{Sequential ignorability}: for each $t$, $A_t$ is independent of all future
counterfactual covariates and outcomes given $H_t$. (iii) \emph{Positivity}: $\mu_t(a\mid H_t)\ge
c_\mu>0$ almost surely for all $a\in\Aset$ and $t$, and $p_g\ge c_p>0$ for all $g$.
\end{assumption}

Small logging probabilities inflate the variance of every estimator through the weights $\omega_t$
defined below.

For $\theta\in\Theta$ define the cumulative importance weights
$\omega_t(\theta)=\prod_{s=1}^t\pi_{\theta,s}(A_s\mid H_s)/\mu_s(A_s\mid H_s)$, $\omega_0=1$, and the
$Q$-functions of outcome $R^{(m)}$ under $\pi_\theta$ by backward recursion,
\begin{gather*}
V^{(m)}_{T+1,\theta}(H_{T+1})=R^{(m)},\qquad
Q^{(m)}_{t,\theta}(h_t,a_t)=\E\{V^{(m)}_{t+1,\theta}(H_{t+1})\mid H_t=h_t,A_t=a_t\},\\
V^{(m)}_{t,\theta}(h_t)=\textstyle\sum_{a\in\Aset}\pi_{\theta,t}(a\mid h_t)\,Q^{(m)}_{t,\theta}(h_t,a),\qquad t=T,\ldots,1 .
\end{gather*}

\begin{lemma}[Identification; \citep{robins1986new}]\label{lem:ident}
Under Assumption~\ref{ass:causal}, $V_g^{(m)}(\theta)=\E\{\omega_T(\theta)R^{(m)}\mid G=g\}
=\E\{V^{(m)}_{1,\theta}(H_1)\mid G=g\}$ for every $\theta$, $g$ and $m$. More generally,
$\E\{f(H^{\pi_\theta}_{T+1})\}=\E\{\omega_T(\theta)f(H_{T+1})\}$ for every bounded $f$.
\end{lemma}

The statistical model $\cM$, which we call the nonparametric model, consists of all laws of $O$ that
satisfy Assumption~\ref{ass:causal}(iii) for some constants $c_\mu,c_p>0$. For every $P\in\cM$ we define $V_g^{(m)}(\theta;P)$ as the identified
functional $\E_P\{\omega^P_T(\theta)R^{(m)}\mid G=g\}$, where $\omega^P_T$ uses the logging policy of
$P$, and $C_j(\vth;P)$ and $\Vstar(P)$ are defined through it; under Assumption~\ref{ass:causal} this
coincides with the counterfactual definition at $P_0$. Writing $\mu_P$ for the logging policy of $P$, we
denote by $\cM_\mu=\{P\in\cM:\mu_P=\mu\}$ the submodel in which the logging policy is known.

Define the doubly robust pseudo-outcome, with nuisance parameters $\eta=(\mu,Q)$,
\begin{equation}\label{eq:psi}
\psi^{(m)}_\theta(O;\eta)=V^{(m)}_{1,\theta}(H_1)+\sum_{t=1}^T\omega_t(\theta)\bigl\{V^{(m)}_{t+1,\theta}(H_{t+1})-Q^{(m)}_{t,\theta}(H_t,A_t)\bigr\}.
\end{equation}

\begin{lemma}[Fixed-policy EIF]\label{lem:eif-fixed}
Under Assumption~\ref{ass:causal}, for fixed $\theta$ the functional $V_g^{(m)}(\theta)$ is pathwise
differentiable in the nonparametric model, with efficient influence function
\[
\varphi^{(m)}_{g,\theta}(O)=\frac{\1\{G=g\}}{p_g}\bigl\{\psi^{(m)}_\theta(O;\eta)-V^{(m)}_g(\theta)\bigr\},
\qquad
\varphi^{(m)}_{+,\theta}(O)=\psi^{(m)}_\theta(O;\eta)-V_+^{(m)}(\theta).
\]
The same functions are the EIFs in the model in which $\mu$ is known. The proportions have EIF
$\1\{G=g\}-p_g$ in both models. The status-quo means have EIF $\1\{G=g\}(R-s_g)/p_g$ in the
nonparametric model and $\1\{G=g\}(\psi^{(0)}_\mu-s_g)/p_g$ when $\mu$ is known, where $\psi^{(0)}_\mu$
is \eqref{eq:psi} with $\pi=\mu$ (so that all weights equal one). For any $\bar\eta=(\bar\mu,\bar
Q)$ that is bounded with $\bar\mu_t\ge c>0$, the pseudo-outcome satisfies
\begin{equation}\label{eq:remainder}
\E\{\psi^{(m)}_\theta(O;\bar\eta)\}-V^{(m)}_+(\theta)=\sum_{t=1}^T\E\Bigl[\{\bar\omega_t(\theta)-\omega_t(\theta)\}\{\tilde Q^{(m)}_{t,\theta}-\bar Q^{(m)}_{t,\theta}\}(H_t,A_t)\Bigr],
\end{equation}
where $\bar\omega_t$ uses $\bar\mu$, $\bar V^{(m)}_{t,\theta}(h_t)=\sum_a\pi_{\theta,t}(a\mid h_t)\bar Q^{(m)}_{t,\theta}(h_t,a)$,
$\bar V^{(m)}_{T+1,\theta}=R^{(m)}$, and $\tilde Q^{(m)}_{t,\theta}(h_t,a_t)=\E\{\bar V^{(m)}_{t+1,\theta}(H_{t+1})\mid h_t,a_t\}$.
\end{lemma}

Lemma~\ref{lem:eif-fixed} collects classical results \citep{murphy2001marginal,vanderlaan2003unified,bang2005doubly,tsiatis2019dynamic,kallus2020double};
the exact remainder \eqref{eq:remainder} is derived in Section~\ref{S-sec:proof-lemma2} of the
Supplement. The vector $\bPhi_\theta$ of influence functions of $\bV(\theta)$ and its derivative in
$\theta$ are the building blocks of Section~\ref{sec:theory}. The remainder \eqref{eq:remainder} is stated for
$V_+$; the same identity holds for $\E\{\1(G=g)\psi_\theta(O;\bar\eta)\}/p_g-V_g(\theta)$ with remainder
$p_g^{-1}\sum_t\E[\1\{G=g\}(\bar\omega_t-\omega_t)(\tilde Q_t-\bar Q_t)]$, because $G$ is part of $H_1$. It
is a sum of products of weight errors and one-step regression errors. It vanishes if either $\bar\mu=\mu$ or every regression is correct; in fact it
vanishes if, for some $t_0$, the propensities are correct up to stage $t_0$ and the regressions are
correct after $t_0$. Under the smoothness condition below, $\theta\mapsto\bPhi_\theta(O)$ is
differentiable and we write $\nabla_\theta\bPhi_\theta$ for its derivative.

\section{Efficiency theory for the constrained optimum}\label{sec:theory}

\subsection{Conditions}

Write $\Psi_{j,\vth}(O)=\nabla_{\bv}c_j\{\vth,\bV(\theta)\}\T\bPhi_\theta(O)$ for the EIF of
$C_j(\vth;P)$ at a fixed $\vth$ (by Lemma~\ref{lem:eif-fixed} and the chain rule; here and below
$\bPhi_\theta$ uses the nonparametric EIFs, and Remark~\ref{rem:knownmu} discusses the known-$\mu$
model), and
$L(\vth,\lambda;P)=C_0(\vth;P)+\sum_{j=1}^J\lambda_jC_j(\vth;P)$ for the Lagrangian.

\begin{condition}[Smoothness]\label{cond:smooth}
$\Theta\times\cU$ is compact; for all $t$, $a$ and $h$, $\theta\mapsto\pi_{\theta,t}(a\mid h)$ is three
times continuously differentiable with $\|\nabla^k_\theta\log\pi_{\theta,t}(a\mid h)\|\le M$ for
$k=1,2,3$; $|R^{(m)}|\le M$; and $c_0,\ldots,c_J$ are three times continuously differentiable on an
open set containing $\{(\vth,\bV(\theta;P)):\vth\in\Theta\times\cU\}$ for all $P$ in a neighbourhood of
$P_0$ (for criterion (k) of Table~\ref{tab:criteria} this requires $v_g>0$).
\end{condition}

Boundedness of the outcomes simplifies the proofs and could be replaced by moment conditions. Two
derivatives suffice for Theorem~\ref{thm:eif}; the third is used in Section~\ref{sec:estimation}.

\begin{condition}[Regular constrained optimum]\label{cond:regular}
At $P_0$: (a) problem \eqref{eq:general} has a unique global maximiser $\vthstar$, which lies in
the interior of $\Theta\times\cU$; (b) \emph{linear independence}: the gradients
$\nabla_\vth C_j(\vthstar;P_0)$, $j\in\cA^\star=\{j\ge1:C_j(\vthstar;P_0)=0\}$, are linearly
independent; (c) \emph{strict complementarity}: the (then unique) KKT multipliers satisfy
$\lamstar_j>0$ for $j\in\cA^\star$; (d) \emph{second-order sufficiency}: $w\T\nabla^2_\vth
L(\vthstar,\lamstar;P_0)w<0$ for every $w\neq0$ with $\nabla_\vth C_j(\vthstar;P_0)\T w=0$ for all
$j\in\cA^\star$.
\end{condition}

Conditions~\ref{cond:regular}(b)--(d) are the standard conditions of nonlinear sensitivity analysis
\citep{fiacco1983introduction,bonnans2000perturbation,nocedal2006numerical}. For the running
example, (b) requires $\nabla_\theta\{V_A-V_B\}(\thstar)\neq0$ when the constraint binds; (c)
excludes the knife-edge case in which the unconstrained optimum lies exactly on the constraint
boundary; and (d) holds when $\thstar$ is a nondegenerate local maximum along the constraint
surface. Under (b) and (d) the KKT matrix
\[
\Kstar=\begin{pmatrix}\nabla^2_\vth L(\vthstar,\lamstar;P_0)&\nabla_\vth C_{\cA^\star}(\vthstar;P_0)\\
\nabla_\vth C_{\cA^\star}(\vthstar;P_0)\T&0\end{pmatrix}
\]
is nonsingular (Lemma~\ref{S-lem:kkt-nonsingular}), where $\nabla_\vth C_{\cA^\star}$ has columns
$\nabla_\vth C_j$, $j\in\cA^\star$. All conditions can be checked numerically on estimated
quantities.

\subsection{Pathwise differentiability of the constrained optimal value}

\begin{theorem}\label{thm:eif}
Suppose Assumption~\ref{ass:causal} and Conditions~\ref{cond:smooth}--\ref{cond:regular} hold.
\begin{enumerate}[label=(\roman*),leftmargin=*]
\item $P\mapsto\Vstar(P)$ is pathwise differentiable at $P_0$ in the nonparametric model, with
efficient influence function
\begin{equation}\label{eq:eif}
\begin{aligned}
\phistar(O)&=\Psi_{0,\vthstar}(O)+\sum_{j\in\cA^\star}\lamstar_j\Psi_{j,\vthstar}(O)\\
&=\sum_{j\in\{0\}\cup\cA^\star}\lamstar_j\,\nabla_{\bv}c_j\{\vthstar,\bV(\thstar)\}\T\bPhi_{\thstar}(O),
\qquad\lamstar_0=1 .
\end{aligned}
\end{equation}
The semiparametric efficiency bound for $\Vstar$ is $\E\{\phistar(O)^2\}$.
\item The optimiser and the active multipliers are pathwise differentiable, with joint influence function
\begin{equation}\label{eq:if-theta-lambda}
\begin{pmatrix}\mathrm{IF}_{\vth}(O)\\ \mathrm{IF}_{\lambda}(O)\end{pmatrix}
=-(\Kstar)^{-1}\begin{pmatrix}\sum_{j\in\{0\}\cup\cA^\star}\lamstar_j\nabla_\vth\Psi_{j,\vth}(O)\big|_{\vth=\vthstar}\\[2pt]
\bigl(\Psi_{j,\vthstar}(O)\bigr)_{j\in\cA^\star}\end{pmatrix}.
\end{equation}
Here $\nabla_\vth\Psi_{j,\vth}$ is the total derivative of
$\vth\mapsto\nabla_{\bv}c_j\{\vth,\bV(\theta)\}\T\bPhi_\theta$, including the derivative of the coefficient
$\nabla_{\bv}c_j\{\vth,\bV(\theta)\}$.
\end{enumerate}
\end{theorem}

For the running example, \eqref{eq:eif} reads
\begin{equation}\label{eq:eif-example}
\phistar=\varphi_{+,\thstar}+\lamstar\bigl(\varphi_{A,\thstar}-\varphi_{B,\thstar}\bigr),
\end{equation}
with $\lamstar=0$ when the constraint is slack. A perturbation that raises $V_A(\thstar)$ relaxes the
binding constraint and so raises $\Vstar$. The multiplier is the shadow price of fairness:
$\partial\Vstar(\delta)/\partial\delta=\lamstar(\delta)$ (by the argument of the proof below with $\delta$
in place of $\varepsilon$; see also \citep{fiacco1983introduction}), so relaxing the tolerance from
$\delta$ to $\delta+\Delta$ lowers the price of fairness by about $\lamstar\Delta$.
Section~\ref{S-sec:remarks} of the Supplement compares the efficiency bounds of the constrained and
unconstrained optimal values.

\begin{proof}[Outline of proof]
Let $P_\varepsilon$ be a submodel with $dP_\varepsilon/dP_0=1+\varepsilon s$ for a bounded score
$s$. The map $(\vth,\varepsilon)\mapsto C_j(\vth;P_\varepsilon)$ is three times continuously differentiable,
with $\partial_\varepsilon C_j(\vth;P_\varepsilon)|_{\varepsilon=0}=\E_0(\Psi_{j,\vth}s)$
(Lemma~\ref{S-lem:smooth-submodel}). Because $\Kstar$ is nonsingular, the implicit function theorem
applied to the KKT system
\[
F(\vth,\lambda_{\cA^\star},\varepsilon)=\begin{pmatrix}\nabla_\vth C_0(\vth;P_\varepsilon)+\sum_{j\in\cA^\star}\lambda_j\nabla_\vth C_j(\vth;P_\varepsilon)\\
\bigl(C_j(\vth;P_\varepsilon)\bigr)_{j\in\cA^\star}\end{pmatrix}=0
\]
yields a continuously differentiable path through $(\vthstar,\lamstar)$, denoted
$\{\vth(\varepsilon),\lambda(\varepsilon)\}$, whose derivative at $\varepsilon=0$ is
$-(\Kstar)^{-1}\partial_\varepsilon F$. Strict
complementarity, second-order sufficiency, linear independence at nearby points (by continuity) and
uniqueness of the global maximiser imply that
$\vth(\varepsilon)$ is the global solution of the perturbed problem for small $|\varepsilon|$, with the
same active set (Supplement, Section~\ref{S-sec:proof-thm1}); this gives part (ii). Differentiating
$\Vstar(P_\varepsilon)=C_0\{\vth(\varepsilon);P_\varepsilon\}$,
\[
\begin{aligned}
\frac{d\Vstar(P_\varepsilon)}{d\varepsilon}\Big|_0&=\nabla_\vth C_0\T\dot\vth+\E_0(\Psi_{0,\vthstar}s)
=-\sum_{j\in\cA^\star}\lamstar_j\nabla_\vth C_j\T\dot\vth+\E_0(\Psi_{0,\vthstar}s)\\
&=\E_0\Bigl\{\Bigl(\Psi_{0,\vthstar}+\sum_{j\in\cA^\star}\lamstar_j\Psi_{j,\vthstar}\Bigr)s\Bigr\},
\end{aligned}
\]
where the second equality is KKT stationarity and the third follows from differentiating the
identities $C_j\{\vth(\varepsilon);P_\varepsilon\}\equiv0$, $j\in\cA^\star$. Equivalently,
$\Vstar(P_\varepsilon)=L\{\vth(\varepsilon),\lambda(\varepsilon);P_\varepsilon\}$ and
$\partial_\lambda L=(C_j)_{j\in\cA^\star}$ vanishes along the path, so the derivative of the multiplier
does not enter. Bounded scores are dense in the tangent space, so
$\phistar$ is the canonical gradient.
\end{proof}

\begin{remark}[Known logging policy]\label{rem:knownmu}
If $\mu$ is known, the tangent space is smaller, but $\phistar$ remains the efficient influence
function provided the criterion does not involve the status-quo means $s_g$, because each component
of $\bPhi_\theta$ for the counterfactual values is orthogonal to the scores of the logging policy
(Section~\ref{S-sec:proof-lemma2}). If the criterion involves $s_g$, as in Table~\ref{tab:criteria}(g), the
same formula holds with $\varphi^s_g$ replaced by its known-$\mu$ version from
Lemma~\ref{lem:eif-fixed} (this is the projection of $\phistar$ onto the smaller tangent space); the
sample mean of $R$ in group $g$ is then consistent and asymptotically normal but, in general, not
efficient.
\end{remark}

\subsection{Other functionals of the constrained optimum}

\begin{corollary}\label{cor:other}
Under the conditions of Theorem~\ref{thm:eif}:
\begin{enumerate}[label=(\alph*),leftmargin=*]
\item If the unconstrained problem also has a unique global maximiser $\theta_{\rm unc}$, in the interior
of $\Theta$, with negative definite Hessian, then $\PoF=V^\star_{\rm unc}-\Vstar$ is pathwise differentiable with EIF
$\varphi_{+,\theta_{\rm unc}}-\phistar$.
\item For a continuously differentiable map $\chi$, the functional
$\chi^\star=\chi\{\vthstar,\bV(\thstar)\}$ has EIF
$\nabla_{\bv}\chi\T\bPhi_{\thstar}+\{\nabla_\vth\chi+\nabla_\vth\bV(\thstar)\T\nabla_{\bv}\chi\}\T\mathrm{IF}_\vth$,
with $\mathrm{IF}_\vth$ from \eqref{eq:if-theta-lambda} and $\nabla_\vth\bV(\thstar)$ the $q\times(d+d_u)$
Jacobian (with zero columns for $u$). In particular, the value delivered to group
$g$ by the fair policy, $V_g(\thstar)$, has EIF
$\varphi_{g,\thstar}+\nabla_\theta V_g(\thstar)\T\mathrm{IF}_\theta$.
\item If $j\in\cA^\star$, the realised constraint value $C_j\{\vthstar(P_\varepsilon);P_\varepsilon\}$ is
identically zero for $|\varepsilon|$ small along every submodel, so its EIF is zero.
\end{enumerate}
\end{corollary}

Unlike $\Vstar$, the group values $V_g(\thstar)$ in (b) depend on how the fair policy moves when the
distribution changes; this sensitivity is the $\vth$-block of $-(\Kstar)^{-1}\partial_\varepsilon F$ and
reflects the curvature of the Lagrangian and of the active constraints. Treating $\thstar$ as fixed and using $\varphi_{g,\thstar}$ alone gives incorrect
standard errors (Section~\ref{sec:sim}).

\subsection{Nonregular configurations}\label{sec:nonregular}

\begin{remark}[Nonregular cases]\label{rem:nonregular}
Condition~\ref{cond:regular} can fail in two main ways. (i) If the maximiser or the multipliers are
not unique, $\Vstar$ is typically only directionally differentiable; under additional conditions its
derivative along submodels has the form $\max_{\vth\in S^\star}\min_{\lambda\in\Lambda(\vth)}\E_0\{(\Psi_{0,\vth}+\sum_j
\lambda_j\Psi_{j,\vth})s\}$ over the solution set $S^\star$ and the multiplier sets $\Lambda(\vth)$
\citep{shapiro1991asymptotic,bonnans2000perturbation}. This is nonlinear in $s$, so no regular
estimator exists \citep{hirano2012impossibility}; pointwise valid inference can use the methods of
\citet{fang2019inference} or the $m$-out-of-$n$ bootstrap \citep{chakraborty2014inference}.
(ii) If the unconstrained optimum lies exactly on the constraint boundary ($\lamstar=0$ with an active
constraint) and $\nabla^2_\theta V(\theta_{\rm unc})$ is negative definite, $\Vstar$ remains pathwise
differentiable with EIF $\varphi_{+,\thstar}$, because the constrained optimum moves by $O(\varepsilon)$
away from a point at which $\nabla_\theta V=0$. In contrast, $\thstar$ and $\lamstar$ are only
directionally differentiable, and $\PoF$ has a zero derivative, so its estimator is $O_p(n^{-1})$ with a
non-Gaussian limit and Wald intervals degenerate. Accordingly, the results of
Section~\ref{sec:estimation} are pointwise in $P_0$: as $\lamstar\to0$, intervals for $\PoF$,
$\lamstar$ and the functionals of Corollary~\ref{cor:other}(b) are not uniformly valid. The influence
function \eqref{eq:eif} of $\Vstar$ varies continuously with $\lamstar$, so the interval for $\Vstar$ is
less sensitive to the boundary (Section~\ref{sec:simresults}); its uniform validity is not proved here.
\end{remark}

\section{Estimation and inference}\label{sec:estimation}

\subsection{Cross-fitted doubly robust estimation of the group functionals}

Randomly split $\{1,\ldots,n\}$ into $K_{\rm f}$ folds $I_1,\ldots,I_{K_{\rm f}}$. For each fold $k$, estimate the
nuisances from the other folds: the logging policy $\hat\mu^{(-k)}$ (or use the logged
probabilities), and, for any $\theta$, the $Q$-functions by backward sequential regression:
regress $R^{(m)}$ on $(H_T,A_T)$, and for $t<T$ regress
$\hat V^{(m),(-k)}_{t+1,\theta}(H_{t+1})=\sum_a\pi_{\theta,t+1}(a\mid H_{t+1})\hat Q^{(m),(-k)}_{t+1,\theta}(H_{t+1},a)$
on $(H_t,A_t)$. For $i\in I_k$ let $\hat\psi^{(m)}_\theta(O_i)$ be \eqref{eq:psi} evaluated at
these nuisances, and set
\[
\hV_g^{(m)}(\theta)=\frac{\sum_i\1\{G_i=g\}\hat\psi^{(m)}_\theta(O_i)}{\sum_i\1\{G_i=g\}},\qquad
\hV_+^{(m)}(\theta)=\Pn\hat\psi^{(m)}_\theta,
\]
and the estimated influence functions $\hat\bPhi_\theta(O_i)$ are obtained by substituting these
quantities into Lemma~\ref{lem:eif-fixed}. When the regressions are linear smoothers (series, splines, kernel
ridge), the regression matrices do not depend on $\theta$, so $\hV(\theta)$ for a new $\theta$
costs only matrix--vector products; this makes the optimisation below inexpensive. The normalised
(H\'ajek) inverse probability weighted estimator
\[
\hV_g^{\rm IPW}(\theta)=\frac{\sum_i\1\{G_i=g\}\omega_{iT}(\theta)R_i}{\sum_i\1\{G_i=g\}\omega_{iT}(\theta)},
\]
with $\omega_{iT}$ the weights of individual $i$, is a simpler, less efficient alternative.

\subsection{Algorithm}

Let $\hC_j(\vth)=c_j\{\vth,\hV(\theta)\}$. The estimator of $\Vstar$ is
$\hV^\star=\hC_0(\hvth)$, where $\hvth$ maximises $\hC_0$ subject to $\hC_j\ge0$. The empirical
problem is smooth but in general not concave, so we combine a global search with local sequential
quadratic programming \citep{kraft1988software,nocedal2006numerical}; Algorithm~\ref{alg:main}
gives the details.

\begin{algorithm}[t]
\caption{Estimation of and inference on the fairness-constrained optimal value}\label{alg:main}
\begin{algorithmic}[1]
\Require Data $O_1,\ldots,O_n$; policy class $\{\pi_\theta\}$; criterion $(c_0,\ldots,c_J)$; number of folds $K_{\rm f}$;
grid $\mathcal{N}\subset\Theta\times\cU$; numerical tolerance $\tau$; level $\alpha$.
\State Fit cross-fitted nuisances; precompute regression (smoother) matrices for every fold and stage.
\State \textbf{Global search.} Evaluate $\hC_0,\ldots,\hC_J$ on $\mathcal{N}$; keep the best feasible points.
\State \textbf{Local refinement.} Run SQP from the best feasible grid points and from the unconstrained
solution, with gradients by central differences; set $\hvth$ to the best feasible local solution.
\State \textbf{Multipliers.} Let $\widehat{\cA}=\{j:|\hC_j(\hvth)|\le\tau\}$ and
$\hlam=\arg\min_{\lambda\ge0}\|\nabla\hC_0(\hvth)+\sum_{j\in\widehat\cA}\lambda_j\nabla\hC_j(\hvth)\|^2$ (non-negative least squares);
$\hlam_j=0$ for $j\notin\widehat\cA$.
\State \textbf{Influence functions.} $\hat\varphi^\star(O_i)=\sum_{j\in\{0\}\cup\widehat\cA}\hlam_j\nabla_{\bv}c_j\{\hvth,\hV(\hth)\}\T\hat\bPhi_{\hth}(O_i)$
($\hlam_0=1$); $\hat\sigma^2=\Pn(\hat\varphi^{\star2})$.
\State \textbf{Output.} $\hV^\star\pm z_{1-\alpha/2}\hat\sigma/\sqrt n$. Repeat with $J=0$ to obtain $\hat\theta_{\rm unc}$, and report
$\widehat\PoF=\hC_0(\hat\theta_{\rm unc})-\hV^\star$ with influence function $\hat\varphi_{+,\hat\theta_{\rm unc}}-\hat\varphi^\star$.
\State \textbf{Optional.} Estimate $(\Kstar)^{-1}$ by differencing to obtain $\mathrm{IF}_\vth$, $\mathrm{IF}_\lambda$ and the
Corollary~\ref{cor:other} intervals; compute the certified policy of Proposition~\ref{prop:cert}.
\end{algorithmic}
\end{algorithm}

In our implementation the grid has $11^d$ points, SQP is the SLSQP routine of SciPy with tolerance
$10^{-12}$, central differences use step $10^{-4}$ and $\tau=10^{-6}$. Grid evaluation is cheap
because each point requires only matrix--vector products. Alternatives, such as dual ascent on
$\lambda$ or augmented Lagrangian methods, can be substituted; the theory requires only that
$\hvth$ be a global maximiser of the empirical problem.

\subsection{Asymptotic theory}

\begin{condition}[Uniform doubly robust expansion]\label{cond:est}
(a) $\sup_{\theta\in\Theta}\|\hV(\theta)-\bV(\theta)-\Pn\bPhi_\theta\|=o_p(n^{-1/2})$;
(b) $\sup_\theta\|\nabla_\theta\hV(\theta)-\nabla_\theta\bV(\theta)\|=O_p(n^{-1/2})$ and
$\sup_\theta\|\nabla^2_\theta\hV(\theta)-\nabla^2_\theta\bV(\theta)\|=o_p(1)$;
(c) $\sup_\theta\Pn\|\hat\bPhi_\theta-\bPhi_\theta\|^2=o_p(1)$;
(d) $\hvth$ is a global maximiser of the empirical problem.
\end{condition}

Condition~\ref{cond:est}(a)--(c) holds for the cross-fitted estimator if the nuisance estimators are
consistent uniformly in $\theta$, together with their $\theta$-derivatives, the products of
propensity and regression errors in \eqref{eq:remainder} are $o_p(n^{-1/2})$ uniformly in $\theta$, and
their $\theta$-derivatives are $O_p(n^{-1/2})$; Lemma~\ref{S-lem:sufficient} gives a sketch of the
argument. The $O_p(n^{-1/2})$ rate in Condition~\ref{cond:est}(b) is needed only for part (v) below;
parts (i)--(iv) and Proposition~\ref{prop:cert} require only a gradient error of order $o_p(n^{-1/4})$,
and the Hessian part of (b) is used only for the variance estimator in (v). If $\mu$ is known, as on adaptive platforms that log their
randomisation probabilities, the product condition holds trivially and only consistency of the
regressions is needed. The uniformity over $\theta$ is mild because $\Theta$ is finite-dimensional
and everything is Lipschitz in $\theta$ under Condition~\ref{cond:smooth}.

\begin{theorem}\label{thm:est}
Suppose Assumption~\ref{ass:causal} and Conditions~\ref{cond:smooth}--\ref{cond:est} hold. Then:
\begin{enumerate}[label=(\roman*),leftmargin=*]
\item $\hvth\to\vthstar$ and $\hlam\to\lamstar$ in probability, and $\Pr(\widehat\cA=\cA^\star)\to1$
for any fixed tolerance $0<\tau<\min_{j\notin\cA^\star}C_j(\vthstar;P_0)$ (with $\min\emptyset=+\infty$);
any $\tau=\tau_n\downarrow0$ also works, because active constraints are solved exactly.
\item $\sqrt n(\hV^\star-\Vstar)=n^{-1/2}\sum_{i=1}^n\phistar(O_i)+o_p(1)\rightsquigarrow N\{0,\E(\phistar{}^2)\}$;
hence $\hV^\star$ is regular and asymptotically efficient in the nonparametric model.
\item $\hat\sigma^2\to\E(\phistar{}^2)$ in probability, so if $\E(\phistar{}^2)>0$ the interval in
Algorithm~\ref{alg:main} has asymptotic coverage $1-\alpha$.
\item Under the condition of Corollary~\ref{cor:other}(a), and if $\hat\theta_{\rm unc}$ is a global maximiser
of $\hV_+$ over $\Theta$, $\sqrt n(\widehat\PoF-\PoF)$ is
asymptotically linear with influence function $\varphi_{+,\theta_{\rm unc}}-\phistar$, whose variance is
estimated consistently by $\Pn(\hat\varphi_{+,\hat\theta_{\rm unc}}-\hat\varphi^\star)^2$; the Wald interval is
valid when this variance is positive, which requires the constraint to bind.
\item If, in addition, $\sqrt n\{\nabla_\theta\hV(\theta)-\nabla_\theta\bV(\theta)\}=\Gn\nabla_\theta\bPhi_\theta+o_p(1)$
uniformly in a neighbourhood of $\thstar$, then $(\hvth,\hlam)$ is asymptotically linear with
influence function \eqref{eq:if-theta-lambda}, and the plug-in estimators of the functionals in
Corollary~\ref{cor:other}(b) are asymptotically linear with the stated influence functions. If also
$\sup_\theta\Pn\|\nabla_\theta\hat\bPhi_\theta-\nabla_\theta\bPhi_\theta\|^2=o_p(1)$, the plug-in variance
estimators based on a difference-quotient estimate of $\Kstar$, with step $h_n\to0$ suitably slowly, are
consistent.
\item Suppose instead that $\mu$ is known and used in the weights, and the regressions converge to
limits $\bar Q$ that need not equal $Q$, where each $\bar Q_{t,\theta}$ is bounded and
$\theta\mapsto\bar Q_{t,\theta}(h,a)$ is Lipschitz uniformly in $(h,a)$. If Condition~\ref{cond:est} holds with $\bPhi_\theta$ replaced by
$\bar\bPhi_\theta$, the influence functions of Lemma~\ref{lem:eif-fixed} computed with $\bar Q$ (and, for
status-quo means, the influence function of the sample mean), then (i), (iii), (iv), the expansion in
(ii), and Proposition~\ref{prop:cert} hold with $\phistar$, $\varphi_{+,\theta_{\rm unc}}$ and
$\Psi_{1,\vthstar}$ replaced by $\bar\varphi^\star=\sum_{j}\lamstar_j\nabla_{\bv}c_j\T\bar\bPhi_{\thstar}$,
$\bar\varphi_{+,\theta_{\rm unc}}$ and $\bar\Psi_{1,\vthstar}$. Since $\bar\varphi^\star-\phistar$ lies in the
closed span of the logging-policy scores, $\hV^\star$ is regular in the model $\cM_\mu$ with $\mu$ known,
and its interval is valid. It is efficient in $\cM_\mu$ only if $\bar\varphi^\star$ equals the known-$\mu$
efficient influence function of Remark~\ref{rem:knownmu}; this holds if $\bar Q_{\cdot,\thstar}=Q_{\cdot,\thstar}$
and no $s_g$ enters the criterion, and fails under generic misspecification.
\end{enumerate}
\end{theorem}

The feasible set of the empirical problem is random, yet it causes no first-order bias: because the
estimated constraints are solved exactly at the active set, the estimation error of the constraints
enters $\hV^\star$ only through the term $\sum_j\lamstar_j\Psi_{j,\vthstar}$ of $\phistar$. The
proof (Supplement, Section~\ref{S-sec:proof-thm2}) uses the identity
$\hV^\star=\hC_0(\hvth)+\sum_{j\in\cA^\star}\lamstar_j\hC_j(\hvth)$, valid with probability tending
to one, a second-order expansion of the population Lagrangian at its stationary point $\vthstar$
and stochastic equicontinuity. Only $\|\hvth-\vthstar\|=o_p(n^{-1/4})$ is needed for (ii). Part (vi)
covers the situation of our simulations, where the logged propensities are known and the regressions
are misspecified: by \eqref{eq:remainder} the pseudo-outcome is then unbiased for every $\theta$, so the
proof goes through with $\bar\bPhi$ in place of $\bPhi$ (Supplement, Step 8 of
Section~\ref{S-sec:proof-thm2}). Numerical aspects of Algorithm~\ref{alg:main} are discussed in
Section~\ref{S-sec:remarks} of the Supplement.

\subsection{Feasibility of the estimated policy and certification}\label{sec:inference}

\begin{proposition}\label{prop:cert}
Suppose the conditions of Theorem~\ref{thm:est} hold with a single active constraint,
$\cA^\star=\{1\}$ (other constraints may be present but slack), and
$\sigma_1^2=\Var(\Psi_{1,\vthstar})>0$.
\begin{enumerate}[label=(\alph*),leftmargin=*]
\item $\sqrt n\,C_1(\hvth;P_0)=-n^{-1/2}\sum_i\Psi_{1,\vthstar}(O_i)+o_p(1)$, so
$\Pr\{C_1(\hvth;P_0)\ge0\}\to1/2$; moreover
$C_0(\hvth;P_0)=\Vstar+\lamstar\Pn\Psi_{1,\vthstar}+o_p(n^{-1/2})$.
\item Let $\hvth_\alpha$ be a global maximiser of the empirical problem in which constraint $j=1$ is
replaced by
\[
\hC_1(\vth)-z_{1-\alpha}\hat\sigma_1(\vth)n^{-1/2}\ge0,
\]
where $\hat\Psi_{1,\vth}$ is the estimated influence function of $C_1$ at $\vth$, obtained by
substituting $\hat\bPhi_\theta$, and $\hat\sigma_1^2(\vth)=\Pn\hat\Psi_{1,\vth}^2$ is twice continuously
differentiable, with $\hat\sigma_1$ and its first and second derivatives bounded in probability
uniformly in $\vth$. Then $\Pr\{C_1(\hvth_\alpha;P_0)\ge0\}\to1-\alpha$ and
\[
C_0(\hvth_\alpha;P_0)=\Vstar-\lamstar n^{-1/2}\bigl\{z_{1-\alpha}\sigma_1-\Gn\Psi_{1,\vthstar}\bigr\}+o_p(n^{-1/2}).
\]
\end{enumerate}
\end{proposition}

Part (a) reflects plug-in estimation rather than a defect of a particular estimator: when fairness
binds, the estimated policy lies exactly on the estimated boundary, so sampling error places it on the
wrong side of the population boundary about half of the time, by an amount of order $n^{-1/2}$;
correspondingly its true value can exceed $\Vstar$. Part (b) gives a certified policy whose population fairness holds with
asymptotic probability $1-\alpha$, at a value cost whose first-order mean is
$\lamstar z_{1-\alpha}\sigma_1n^{-1/2}$.
For several active constraints a Bonferroni or simultaneous margin can be used. A related idea in
reinforcement learning is to deploy a learned policy only if a high-confidence bound certifies that
undesirable outcomes, including unfair ones, are unlikely \citep{thomas2019preventing}; such
guarantees hold in finite samples but are typically conservative, whereas the margin in (b) is
asymptotically exact and its cost in value is explicit.

In practice we recommend reporting, for a chosen $\delta$ or a grid of values:
(1) $\hV^\star(\delta)$ with its interval; (2) the one-sided test of $H_0\colon C_1(\theta_{\rm
unc})\ge0$, that the unconstrained optimal policy already satisfies the constraint, based on
$\hC_1(\hat\theta_{\rm unc})$ and the standard error from Corollary~\ref{cor:other}(b) applied to the
unconstrained problem (valid under the condition of Theorem~\ref{thm:est}(v) near $\theta_{\rm unc}$), whose influence
function $\Psi_{1,\theta_{\rm unc}}+\nabla_\theta C_1(\theta_{\rm unc})\T\mathrm{IF}_{\theta_{\rm unc}}$, with
$\mathrm{IF}_{\theta_{\rm unc}}=-\{\nabla^2_\theta V(\theta_{\rm unc})\}^{-1}\nabla_\theta\varphi_{+,\theta_{\rm unc}}$,
accounts for the estimation of $\theta_{\rm unc}$; (3) $\widehat\PoF(\delta)$ with its interval, to be
interpreted with caution unless the test in (2) rejects clearly, because near the boundary the interval
is not reliable (Remark~\ref{rem:nonregular}(ii)); (4) the shadow price $\hlam(\delta)$; (5) the group
values delivered by the fair policy with Corollary~\ref{cor:other} intervals; and (6) the certified
policy of Proposition~\ref{prop:cert}(b).

\section{Numerical studies}\label{sec:sim}

\subsection{Exact verification of the influence functions}\label{sec:exact}

To check Theorem~\ref{thm:eif} and Corollary~\ref{cor:other} independently of any estimation error,
we used a two-stage law with binary group, covariates, actions and outcome, so that all
functionals can be computed exactly (details in Section~\ref{S-sec:exact} of the Supplement). The
policy class is \eqref{eq:softthreshold} with $\kappa=2$ and $\theta\in\R^2$, and $\delta$ was set so
that the gap constraint binds with \input{tables/tex_exact_lambda.tex}. For 20 random bounded
directions $s$ we compared the central finite-difference derivative of each functional along
$dP_\varepsilon=(1+\varepsilon s)dP_0$, $\varepsilon=10^{-5}$, with $\E_0\{\mathrm{IF}(O)s(O)\}$, solving
every perturbed problem by Newton's method on the KKT system to machine precision.
Table~\ref{tab:exact} shows that the formulas of Theorem~\ref{thm:eif} and Corollary~\ref{cor:other}
agree with the numerical derivatives to within $3\times10^{-9}$. In contrast, two natural shortcuts
are off by amounts of the same order as the derivatives themselves: ignoring the constraint, which
drops the multiplier term from the influence function of $\Vstar$, and treating $\thstar$ as fixed
in the influence function of $V_A(\thstar)$. A second check on a three-group law with three policy
parameters, covering an inequality-averse objective with nonlinear constraints and max--min welfare
with two active constraints, gives agreement to within $2\times10^{-8}$
(Table~\ref{S-tab:exact-general} of the Supplement).

\begin{table}[t]
\centering
\caption{Exact check of the influence functions: maximum over 20 random directions $s$ of
$|\text{numerical derivative}-\E_0(\mathrm{IF}\,s)|$, and of the numerical derivative itself.}
\label{tab:exact}
\input{tables/tex_exact.tex}
\end{table}

\subsection{Monte Carlo design}\label{sec:mcdesign}

We simulated $T=3$ tutoring sessions. Group $A$ (the protected group, $\Pr(G=A)=0.4$) starts with
lower mastery, $X_1\sim N(-0.3\,\1\{G=A\},1)$. At session $t$ the platform chooses $A_t=1$ (advance)
or $A_t=0$ (review) with known logged probability
$\mu_t(1\mid H_t)=\expit(0.3+0.6X_t-0.5\,\1\{G=A\}-0.4A_{t-1})$. The session gain is
$Y_t=0.2X_t+A_t(0.3+0.5X_t+0.2X_{t-1}-0.8\,\1\{G=A\})+\epsilon_t$, mastery evolves as
$X_{t+1}=0.6X_t+0.2X_{t-1}+0.5Y_t+\nu_{t+1}$, and the outcome is end-of-course mastery $R=X_4$, with
$\epsilon_t,\nu_t\sim N(0,0.5^2)$ and $X_0=A_0=0$. The dynamics are second-order (non-Markov) and the
benefit of advancing is smaller in group $A$, so the optimal common policy trades off the two
groups. The policy class is \eqref{eq:softthreshold}, which does not use $G$, with $\kappa=2$,
$S_1=X_1$, $S_t=(X_t+X_{t-1})/2$ for $t\ge2$, a threshold $\theta_1$ for the first session and
$\theta_2$ for the later ones, and $\Theta=[-2.5,2.5]^2$. The estimand is \eqref{eq:intro} for
$\delta\in\{0.5,0.6,0.7,0.8,1.0\}$. True values were computed by enumerating all $2^T$ counterfactual
action paths for $10^6$ simulated individuals with common random numbers and solving the KKT system
by Newton's method; they are listed in Table~\ref{S-tab:truth} of the Supplement. The constraint binds for
$\delta\le0.8$, with $\lamstar$ decreasing from $0.41$ to $0.035$, and is slack for $\delta=1$;
Conditions~\ref{cond:regular}(a)--(d) were verified numerically (Section~\ref{S-sec:sim-details}).
The case $\delta=0.8$, with $\lamstar=0.035$, is close to the nonregular configuration of
Remark~\ref{rem:nonregular}(ii).

The doubly robust estimator used $K_{\rm f}=5$ folds, the logged propensities and series regressions on
$(1,G,X_t,X_t^2,GX_t,X_{t-1},X_tX_{t-1},A_{t-1})$ fully interacted with $A_t$; these regressions are
misspecified for $t<T$, because the true $Q$-functions are nonlinear. The H\'ajek IPW estimator
used the same optimisation. Algorithm~\ref{alg:main} was run as described in
Section~\ref{sec:estimation}. In this main scenario, S1, we used $n\in\{1000,2000,4000\}$ and 500
replications; it is covered by Theorem~\ref{thm:est}(vi), because the propensities are known and the
regressions converge to well-defined limits. Three additional scenarios at $n=2000$ (300 replications)
estimate $\mu$ by cross-fitted logistic regression: S2 uses a correctly specified propensity model and
the series regressions; S3 uses a correct propensity model and a constant outcome model; S4 uses the
series regressions and a misspecified propensity model that omits $X_t$ and $A_{t-1}$. With estimated
propensities and regressions that are at best approximately correct, the rate condition of
Theorem~\ref{thm:est} is not guaranteed, so S2--S4 serve as robustness checks. The Gaussian design
does not satisfy the boundedness conditions exactly ($R$ is unbounded and the propensities are bounded
away from zero only with high probability), but in a sample of $10^5$ individuals the logged
propensities lie in $[0.02,0.97]$ (Section~\ref{S-sec:sim-details}). True constraint values and values of estimated
policies are computed on the same counterfactual sample that defines the truth (Monte Carlo standard
error 0.0013 for values and 0.0025 for the gap; Section~\ref{S-sec:sim-details}).

\subsection{Results}\label{sec:simresults}

\input{tables/tex_sim_main.tex}

\input{tables/tex_sim_text.tex}

\section{A synthetic intelligent-tutoring case study}\label{sec:case}

\input{tables/tex_case.tex}

\section{Discussion}\label{sec:discussion}

We have shown that the optimal value of a fairness-constrained policy problem is a regular
statistical functional under standard second-order conditions, and that its efficient influence
function is the Lagrangian combination of the known influence functions of the group values. The
practical consequence is simple: once the constrained optimum and its multipliers have been
computed, efficient inference for the constrained value and the price of fairness requires only the
same influence-function evaluations that are used for fixed policies. The multiplier also measures
how quickly the price of fairness falls as the tolerance is relaxed. Quantities that are not optimal
values, such as the value delivered to each group by the fair policy, require the full sensitivity
analysis of the KKT system.

Several limitations deserve emphasis. Our results concern smooth parametric policy classes. For an
unrestricted class the constrained problem is a linear program in the distribution of action paths,
whose solutions randomise on a set of histories determined by the constraints, and pathwise
differentiability then requires margin conditions, as in \citet{luedtke2016resource} for single-stage
resource constraints; extending Theorem~\ref{thm:eif} to unrestricted multi-stage classes is open.
At the boundary between binding and slack constraints, or when optima are not unique, the problem
is nonregular, and near the boundary Wald intervals for the price of fairness and the multiplier can
be inaccurate in finite samples, as the case $\delta=0.8$ in Section~\ref{sec:sim} illustrates;
uniformly valid inference in this regime remains open. In our simulations the plug-in fair policy is feasible slightly less than half
of the time, and at $n=1000$ the tightened problem of the certified policy often has no solution when
the multiplier is large; finite-sample refinements of the certification margin of Proposition~\ref{prop:cert} would be
useful. We have assumed sequential ignorability;
sensitivity analysis for unmeasured confounding \citep{manski2003partial} or proximal methods
\citep{tchetgen2020introduction} could be combined with our approach. We have also taken the
horizon $T$ fixed; long-horizon settings call for Markov or stationarity assumptions as in
\citet{kallus2020double} and \citet{liao2021off}. Finally, the choice of fairness criterion and
tolerance is a normative decision that statistics cannot settle; what our results provide is an
efficient and valid quantification of its consequences.

\bibliographystyle{unsrtnat}
\bibliography{refs}

\clearpage
\setcounter{section}{0}\renewcommand{\thesection}{S\arabic{section}}
\setcounter{equation}{0}\renewcommand{\theequation}{S\arabic{equation}}
\setcounter{table}{0}\renewcommand{\thetable}{S\arabic{table}}
\setcounter{figure}{0}\renewcommand{\thefigure}{S\arabic{figure}}
\setcounter{lemma}{0}\renewcommand{\thelemma}{S\arabic{lemma}}
\setcounter{theorem}{0}\renewcommand{\thetheorem}{S\arabic{theorem}}
\setcounter{remark}{0}\renewcommand{\theremark}{S\arabic{remark}}
\begin{center}
{\Large\bf Supplementary Material}
\end{center}

\noindent
References to equations, theorems and sections without the prefix ``S'' are to the main paper.
Throughout, $\E$ and $\|\cdot\|$ (for functions) refer to $P_0$ and the $L_2(P_0)$ norm, $\Pn$ is the
empirical measure, $\Gn=\sqrt n(\Pn-P_0)$, and $C_j(\vth)=C_j(\vth;P_0)$. We write
$\vth\in\Theta\times\cU$ and, abusing notation, $\bV(\vth)=\bV(\theta)$ and
$\bPhi_\vth=\bPhi_\theta$.

\section{Proof of Lemma~\ref{lem:eif-fixed} (fixed-policy influence function)}\label{S-sec:proof-lemma2}

The lemma is classical \citep{murphy2001marginal,vanderlaan2003unified,bang2005doubly,tsiatis2019dynamic,kallus2020double};
we include a short derivation for completeness and because the exact remainder
\eqref{eq:remainder} is used in Lemma~\ref{S-lem:sufficient}. Fix $\theta$ and $m$ and drop them
from the notation.

\paragraph{Remainder identity.} Let $\bar\eta=(\bar\mu,\bar Q)$ be arbitrary (bounded, with
$\bar\mu_t\ge c>0$), let $\bar\omega_t$ be the weights computed with $\bar\mu$, and define
\[
\bar V_t(h)=\sum_a\pi_t(a\mid h)\bar Q_t(h,a),\qquad \bar V_{T+1}=R,\qquad
\tilde Q_t(h,a)=\E\{\bar V_{t+1}(H_{t+1})\mid H_t=h,A_t=a\}.
\]
Since $\bar\omega_t$ is a function of $(H_t,A_t)$, iterated expectations give
\begin{equation}\label{S-eq:s1}
\E\{\psi(O;\bar\eta)\}=\E\{\bar V_1(H_1)\}+\sum_{t=1}^T\E\bigl[\bar\omega_t\{\tilde Q_t-\bar Q_t\}(H_t,A_t)\bigr].
\end{equation}
For the true weights and any bounded $f$,
\[
\E\{\omega_tf(H_t,A_t)\}=\E\Bigl\{\omega_{t-1}\sum_a\pi_t(a\mid H_t)f(H_t,a)\Bigr\},
\]
because $\E\{\pi_t(A_t\mid H_t)f(H_t,A_t)/\mu_t(A_t\mid H_t)\mid H_t\}=\sum_a\pi_t(a\mid H_t)f(H_t,a)$.
Applying this with $f=\bar Q_t$, and using
$\E\{\omega_t\bar V_{t+1}(H_{t+1})\}=\E\{\omega_t\tilde Q_t(H_t,A_t)\}$,
\[
\begin{aligned}
\sum_{t=1}^T\E\bigl[\omega_t\{\tilde Q_t-\bar Q_t\}(H_t,A_t)\bigr]
&=\sum_{t=1}^T\bigl[\E\{\omega_t\bar V_{t+1}(H_{t+1})\}-\E\{\omega_{t-1}\bar V_t(H_t)\}\bigr]\\
&=\E(\omega_TR)-\E\{\bar V_1(H_1)\}.
\end{aligned}
\]
By Lemma~\ref{lem:ident}, $\E(\omega_TR)=V_+(\theta)$. Subtracting this display from \eqref{S-eq:s1}
gives \eqref{eq:remainder}:
$\E\{\psi(O;\bar\eta)\}-V_+(\theta)=\sum_t\E[(\bar\omega_t-\omega_t)(\tilde Q_t-\bar Q_t)(H_t,A_t)]$.

\paragraph{Influence function.} Let $\bar P\in\cM$ (i.e.\ satisfying positivity), with nuisances
$\bar\eta$. Since $\E_{\bar P}\{\psi(O;\bar\eta)\}=V_+(\theta;\bar P)$, the identity
gives the von Mises expansion
\[
V_+(\theta;\bar P)-V_+(\theta;P_0)=-\E_0\{\psi(O;\bar\eta)-V_+(\theta;\bar P)\}+\mathrm{Rem}(\bar P,P_0),
\]
with $\mathrm{Rem}$ the right-hand side of \eqref{eq:remainder}, a sum of products of differences
between $\bar P$ and $P_0$. Along a submodel $P_\varepsilon$ with bounded score $s$, both factors
of $\mathrm{Rem}(P_\varepsilon,P_0)$ are $O(\varepsilon)$, so $\mathrm{Rem}=O(\varepsilon^2)$. Moreover,
since $\E_{P_\varepsilon}\{\psi(O;\eta_\varepsilon)-V_+(\theta;P_\varepsilon)\}=0$,
\[
-\E_0\{\psi(O;\eta_\varepsilon)-V_+(\theta;P_\varepsilon)\}=\varepsilon\E_0[\{\psi(O;\eta_\varepsilon)-V_+(\theta;P_\varepsilon)\}s],
\]
and $\psi(\cdot;\eta_\varepsilon)\to\psi(\cdot;\eta)$ uniformly as $\varepsilon\to0$ (the nuisances of $P_\varepsilon$
are smooth in $\varepsilon$ with bounded derivatives, by the density-ratio representation of
Lemma~\ref{S-lem:smooth-submodel}). Hence
$dV_+(\theta;P_\varepsilon)/d\varepsilon|_0=\E_0[\{\psi(O;\eta)-V_+(\theta)\}s]$; see
\citet{hines2022demystifying} for this argument in general. Hence $\psi-V_+$ is a gradient, and it is
the canonical gradient in the nonparametric model. For $V_g=\E\{\1(G=g)\psi\}/p_g$ the quotient rule
gives $\varphi_g=\1\{G=g\}(\psi-V_g)/p_g$; the EIFs of $s_g$ and $p_g$ are standard.

\paragraph{Known logging policy.} If $\mu$ is known, the tangent space is the subspace of
$L_2^0(P_0)$ orthogonal to all scores of the conditional laws of $A_t$ given $H_t$, i.e.\ to all
functions $b(H_t,A_t)-\E\{b(H_t,A_t)\mid H_t\}$. With true nuisances,
$\E\{\omega_s(V_{s+1}-Q_s)\mid H_t,A_t\}=0$ for $s\ge t$, and the terms with $s<t$ and
$V_1(H_1)$ are functions of $H_t$; hence $\E(\psi\mid H_t,A_t)$ does not depend on $A_t$ and
$\psi-V_+$ is orthogonal to these scores. It is therefore in the restricted tangent space and remains
the EIF. The same holds for $\varphi_g$.

\section{Two lemmas for Theorem~\ref{thm:eif}}\label{S-sec:lemmas}

\begin{lemma}[Smoothness along submodels]\label{S-lem:smooth-submodel}
Let $s$ be bounded, $\|s\|_\infty\le B$, $E_0s=0$, and $dP_\varepsilon=(1+\varepsilon s)dP_0$ for
$|\varepsilon|<\varepsilon_0=1/(2B)$. Under Assumption~\ref{ass:causal} and
Condition~\ref{cond:smooth}: (i) $P_\varepsilon\in\cM$, with positivity constant $c_\mu/3$ and
$P_\varepsilon(G=g)\ge p_g/2$;
(ii) $(\vth,\varepsilon)\mapsto C_j(\vth;P_\varepsilon)$ is three times continuously differentiable on
$(\Theta\times\cU)\times(-\varepsilon_0,\varepsilon_0)$; (iii)
$\partial_\varepsilon C_j(\vth;P_\varepsilon)|_{\varepsilon=0}=\E_0(\Psi_{j,\vth}s)$ and
$\partial_\varepsilon\nabla_\vth C_j(\vth;P_\varepsilon)|_{\varepsilon=0}=\E_0(\nabla_\vth\Psi_{j,\vth}s)$.
\end{lemma}

\begin{proof}
Let $\bar s_t(h_t)=\E_0(s\mid H_t=h_t)$ and $\bar s'_t(h_t,a_t)=\E_0(s\mid H_t=h_t,A_t=a_t)$, both
bounded by $B$. Under $P_\varepsilon$ the conditional law of the block $(Y_t,X_{t+1})$ given
$(H_t,A_t)$ has density $\rho^\varepsilon_{t+1}=\{1+\varepsilon\bar s_{t+1}(H_{t+1})\}/\{1+\varepsilon\bar s'_t(H_t,A_t)\}$
relative to $P_0$, the law of $H_1$ has density $\rho^\varepsilon_1=1+\varepsilon\bar s_1(H_1)$, and the
conditional law of $A_t$ given $H_t$ has density $\{1+\varepsilon\bar s'_t\}/\{1+\varepsilon\bar s_t\}\in[1/3,3]$;
together with $P_\varepsilon(G=g)=\E_0\{\1(G=g)(1+\varepsilon s)\}\ge p_g/2$ this gives (i). The g-formula integrates $r_m(h_{T+1})$ against $\prod_t\pi_{\theta,t}(a_t\mid h_t)$ and
the covariate and outcome conditionals; writing the latter under $P_\varepsilon$ as $\rho^\varepsilon_{t}$ times
those under $P_0$ and applying the change-of-measure identity
$\int f\prod_t\pi_{\theta,t}\,dP_0(\text{covariate and outcome blocks})=\E_0\{\omega_T(\theta)f(H_{T+1})\}$, which
needs only positivity, to $f=r_m\prod_{t=1}^{T+1}\rho^\varepsilon_t$ yields
\[
V^{(m)}_+(\theta;P_\varepsilon)=\E_0\Bigl\{\omega_T(\theta)R^{(m)}\prod_{t=1}^{T+1}\rho_t^\varepsilon\Bigr\},\qquad
V^{(m)}_g(\theta;P_\varepsilon)=\frac{\E_0\{\1(G=g)\omega_T(\theta)R^{(m)}\prod_t\rho_t^\varepsilon\}}{p_g+\varepsilon\E_0\{\1(G=g)s\}} .
\]
Each $\rho^\varepsilon_t$ is a ratio of affine functions of $\varepsilon$ with bounded coefficients and
denominator at least $1/2$, hence infinitely differentiable in $\varepsilon$ with derivatives bounded
uniformly in $O$. Under Condition~\ref{cond:smooth} and positivity, $\omega_T(\theta)R^{(m)}$ is
three times continuously differentiable in $\theta$ with bounded derivatives. Dominated convergence
gives (ii) for the components of $\bV$, and the chain rule extends it to $C_j$. By
Lemma~\ref{lem:eif-fixed}, $\partial_\varepsilon\bV(\theta;P_\varepsilon)|_0=\E_0(\bPhi_\theta s)$, which
gives the first claim of (iii) by the chain rule. For the second, (ii) allows the derivatives to be
interchanged:
$\partial_\varepsilon\nabla_\vth C_j|_0=\nabla_\vth\E_0(\Psi_{j,\vth}s)=\E_0(\nabla_\vth\Psi_{j,\vth}s)$,
the last step by dominated convergence, since $\nabla_\theta\bPhi_\theta$ is bounded (the
$Q$-functions are smooth in $\theta$ with bounded derivatives under Condition~\ref{cond:smooth}).
\end{proof}

\begin{lemma}[Nonsingularity of the KKT matrix]\label{S-lem:kkt-nonsingular}
Let $\mathbf{M}$ be a symmetric $p\times p$ matrix and $\mathbf{D}$ a $p\times m$ matrix of rank $m$, with
$w\T\mathbf{M}w<0$ for all $w\neq0$ such that $\mathbf{D}\T w=0$. Then
$\begin{pmatrix}\mathbf{M}&\mathbf{D}\\\mathbf{D}\T&0\end{pmatrix}$ is nonsingular.
\end{lemma}

\begin{proof}
If $\mathbf{M}x+\mathbf{D}y=0$ and $\mathbf{D}\T x=0$, then $x\T\mathbf{M}x=-x\T\mathbf{D}y=-(\mathbf{D}\T x)\T y=0$, so
$x=0$ by the assumption; then $\mathbf{D}y=0$ and $y=0$ because $\mathbf{D}$ has full column rank.
\end{proof}

\section{Proof of Theorem~\ref{thm:eif}}\label{S-sec:proof-thm1}

Fix a bounded score $s$ and the submodel $P_\varepsilon$ of Lemma~\ref{S-lem:smooth-submodel}. Write
$C_j(\vth;\varepsilon)=C_j(\vth;P_\varepsilon)$, $m=|\cA^\star|$, and
$\mathbf{D}(\vth;\varepsilon)$ for the $(d+d_u)\times m$ matrix with columns $\nabla_\vth C_j(\vth;\varepsilon)$,
$j\in\cA^\star$.

\paragraph{Step 1 (implicit function theorem).} For $\lambda\in\R^m$ define
\[
F(\vth,\lambda,\varepsilon)=\begin{pmatrix}\nabla_\vth C_0(\vth;\varepsilon)+\mathbf{D}(\vth;\varepsilon)\lambda\\
(C_j(\vth;\varepsilon))_{j\in\cA^\star}\end{pmatrix}.
\]
By Lemma~\ref{S-lem:smooth-submodel},
$F$ is continuously differentiable near $(\vthstar,\lamstar_{\cA^\star},0)$; $F(\vthstar,\lamstar_{\cA^\star},0)=0$
by the KKT conditions; and $\partial F/\partial(\vth,\lambda)=\Kstar$ at that point, which is
nonsingular by Lemma~\ref{S-lem:kkt-nonsingular} and Condition~\ref{cond:regular}(b),(d). Hence there
are neighbourhoods $\mathcal{B}$ of $(\vthstar,\lamstar_{\cA^\star})$ and $(-\varepsilon_1,\varepsilon_1)$ of
$0$ and a unique continuously differentiable map $\varepsilon\mapsto\{\vth(\varepsilon),\lambda(\varepsilon)\}\in\mathcal{B}$
with $F\{\vth(\varepsilon),\lambda(\varepsilon),\varepsilon\}=0$ and
\begin{equation}\label{S-eq:ift}
\begin{pmatrix}\dot\vth\\ \dot\lambda\end{pmatrix}=-(\Kstar)^{-1}\partial_\varepsilon F(\vthstar,\lamstar_{\cA^\star},0)
=-(\Kstar)^{-1}\E_0\left\{\begin{pmatrix}\nabla_\vth\Psi_{0,\vthstar}+\sum_{j\in\cA^\star}\lamstar_j\nabla_\vth\Psi_{j,\vthstar}\\
(\Psi_{j,\vthstar})_{j\in\cA^\star}\end{pmatrix}s\right\},
\end{equation}
by Lemma~\ref{S-lem:smooth-submodel}(iii). This is part (ii) of the theorem, once Step 2 shows that
$\{\vth(\varepsilon),\lambda(\varepsilon)\}$ are the optimiser and multipliers of the perturbed problem.

\paragraph{Step 2 ($\vth(\varepsilon)$ solves the perturbed problem).}
(a) \emph{Local optimality.} Extend $\lambda(\varepsilon)$ by zeros for $j\notin\cA^\star$. By continuity, for
$|\varepsilon|$ small: $\lambda_j(\varepsilon)>0$ for $j\in\cA^\star$ (Condition~\ref{cond:regular}(c));
$C_j\{\vth(\varepsilon);\varepsilon\}>0$ for $j\notin\cA^\star$; $\mathbf{D}\{\vth(\varepsilon);\varepsilon\}$ has full
column rank; and the second-order condition holds at $\{\vth(\varepsilon),\lambda(\varepsilon)\}$, because
with $Z(\varepsilon)$ a basis of the null space of $\mathbf{D}\{\vth(\varepsilon);\varepsilon\}\T$, chosen
continuously, $Z\T\nabla^2_\vth LZ$ is negative definite at $\varepsilon=0$ and hence nearby. So
$\vth(\varepsilon)$ is a KKT point satisfying the second-order sufficient conditions, and therefore a
strict local maximiser of the perturbed problem \citep[Theorem 12.6]{nocedal2006numerical}.

(b) \emph{Global optimality.} The feasible set of the perturbed problem is compact and contains
$\vth(\varepsilon)$, so a global maximiser $\tilde\vth_\varepsilon$ exists. By
Lemma~\ref{S-lem:smooth-submodel}, $\sup_\vth|C_j(\vth;\varepsilon)-C_j(\vth;0)|\to0$ as
$\varepsilon\to0$. Let $\bar\vth$ be a limit point of $\tilde\vth_\varepsilon$ along
$\varepsilon_k\to0$. Then $C_j(\bar\vth;0)=\lim C_j(\tilde\vth_{\varepsilon_k};\varepsilon_k)\ge0$
for all $j$, and $C_0(\bar\vth;0)=\lim C_0(\tilde\vth_{\varepsilon_k};\varepsilon_k)\ge\lim
C_0\{\vth(\varepsilon_k);\varepsilon_k\}=\Vstar$; by uniqueness (Condition~\ref{cond:regular}(a))
$\bar\vth=\vthstar$. Hence $\tilde\vth_\varepsilon\to\vthstar$. For small $|\varepsilon|$,
$\tilde\vth_\varepsilon$ is interior, the constraints $j\notin\cA^\star$ are inactive at it, and the
gradients of the constraints in $\cA^\star$ are linearly independent there, so $\tilde\vth_\varepsilon$
satisfies the KKT conditions with unique multipliers
$\tilde\lambda_\varepsilon=-(\tilde{\mathbf{D}}\T\tilde{\mathbf{D}})^{-1}\tilde{\mathbf{D}}\T\nabla_\vth C_0(\tilde\vth_\varepsilon;\varepsilon)$
on $\cA^\star$, where $\tilde{\mathbf{D}}=\mathbf{D}(\tilde\vth_\varepsilon;\varepsilon)$ (inactive constraints have zero
multipliers). By continuity $\tilde\lambda_\varepsilon\to\lamstar_{\cA^\star}$, so
$\tilde\lambda_{\varepsilon,j}>0$ and, by complementary slackness, $C_j(\tilde\vth_\varepsilon;\varepsilon)=0$
for all $j\in\cA^\star$. Thus $(\tilde\vth_\varepsilon,\tilde\lambda_\varepsilon)\in\mathcal{B}$ solves
$F=0$, and uniqueness in Step 1 gives $\tilde\vth_\varepsilon=\vth(\varepsilon)$. Consequently
$\Vstar(P_\varepsilon)=C_0\{\vth(\varepsilon);\varepsilon\}$ for $|\varepsilon|$ small, the active set
is constant, and $\lambda(\varepsilon)$ are the multipliers.

\paragraph{Step 3 (derivative of the optimal value).} By the chain rule and
Lemma~\ref{S-lem:smooth-submodel},
\[
\frac{d}{d\varepsilon}\Vstar(P_\varepsilon)\Big|_0=\nabla_\vth C_0(\vthstar;0)\T\dot\vth+\E_0(\Psi_{0,\vthstar}s).
\]
KKT stationarity gives $\nabla_\vth C_0(\vthstar;0)=-\sum_{j\in\cA^\star}\lamstar_j\nabla_\vth C_j(\vthstar;0)$,
and differentiating the identities $C_j\{\vth(\varepsilon);\varepsilon\}\equiv0$, $j\in\cA^\star$, gives
$\nabla_\vth C_j(\vthstar;0)\T\dot\vth=-\E_0(\Psi_{j,\vthstar}s)$. Therefore
\begin{equation}\label{S-eq:deriv}
\frac{d}{d\varepsilon}\Vstar(P_\varepsilon)\Big|_0=\E_0\Bigl\{\Bigl(\Psi_{0,\vthstar}+\sum_{j\in\cA^\star}\lamstar_j\Psi_{j,\vthstar}\Bigr)s\Bigr\}=\E_0(\phistar s).
\end{equation}
Equivalently, $\Vstar(P_\varepsilon)=L\{\vth(\varepsilon),\lambda(\varepsilon);P_\varepsilon\}$, whose
total derivative is $\partial_\vth L\,\dot\vth+\partial_\lambda L\,\dot\lambda+\partial_\varepsilon L$
with $\partial_\vth L=0$ (stationarity) and $\partial_\lambda L=(C_j)_{j\in\cA^\star}=0$ (active
constraints), so the derivative $\dot\lambda$ of the multiplier, given by \eqref{S-eq:ift}, does not
enter \eqref{S-eq:deriv}.

\paragraph{Step 4 (efficient influence function).} Each component of $\bPhi_\theta$ has mean zero, so
$\phistar\in L_2^0(P_0)$, and \eqref{S-eq:deriv} holds for every bounded score. Bounded scores form a
tangent set whose closed linear span is $L^0_2(P_0)$, and $s\mapsto\E_0(\phistar s)$ is continuous,
so $\Vstar$ is pathwise differentiable with canonical gradient $\phistar$
\citep[Section 25.3]{vandervaart1998}. In the model with $\mu$ known, the submodels are those whose
scores satisfy $\E_0(s\mid H_t,A_t)=\E_0(s\mid H_t)$ for all $t$. Bounded scores of this kind are dense in
the restricted tangent space (truncate each covariate and outcome block component
$b(H_{t+1})-\E_0\{b(H_{t+1})\mid H_t,A_t\}$ and re-centre), and \eqref{S-eq:deriv} holds along them, so
$\phistar$ is a gradient in that model. If the criterion does not involve the status-quo means $s_g$,
$\phistar$ is a linear combination of components of $\bPhi_{\thstar}$ that are orthogonal to the
logging-policy scores (Section~\ref{S-sec:proof-lemma2}), so it lies in the closure of the restricted
tangent space and is the canonical gradient there as well. Otherwise the canonical gradient is the
$L_2(P_0)$-projection of $\phistar$ onto that closure. Since all other components of $\bPhi_{\thstar}$
already lie in it, the projection replaces $\varphi^s_g=\1\{G=g\}(R-s_g)/p_g$ by
$\1\{G=g\}(\psi^{(0)}_\mu-s_g)/p_g$. The difference between the two is
$-\1\{G=g\}p_g^{-1}\sum_t\{Q^\mu_t(H_t,A_t)-V^\mu_t(H_t)\}$, where $Q^\mu_t,V^\mu_t$ are the $Q$- and
$V$-functions of $R$ under $\mu$, and it lies in the span of the action scores. This proves
Remark~\ref{rem:knownmu}. The efficiency bound $\E(\phistar{}^2)$ follows from the convolution theorem \citep[Chapter
25]{vandervaart1998}. Part (ii) follows from \eqref{S-eq:ift} and Step 2. \qed

\paragraph{Proof of Corollary~\ref{cor:other}.} (a) The unconstrained problem is the case $J=0$
of Theorem~\ref{thm:eif} (the second-order condition is negative definiteness of the Hessian), so
$V^\star_{\rm unc}$ has EIF $\varphi_{+,\theta_{\rm unc}}$, and the EIF of the difference is the
difference of the EIFs. (b) By the chain rule along the path of Step~1,
$d\chi\{\vth(\varepsilon),\bV(\theta(\varepsilon);P_\varepsilon)\}/d\varepsilon|_0
=\nabla_\vth\chi\T\dot\vth+\nabla_{\bv}\chi\T\{\nabla_\vth\bV(\thstar)\dot\vth+\E_0(\bPhi_{\thstar}s)\}$,
and $\dot\vth=\E_0(\mathrm{IF}_\vth s)$ by \eqref{S-eq:ift}. (c) By Step 2, $C_j\{\vth(\varepsilon);\varepsilon\}\equiv0$.
\qed

\section{Sufficient conditions for Condition~\ref{cond:est}}\label{S-sec:sufficient}

For fold $k$, let $\hat\eta^{(-k)}=(\hat\mu^{(-k)},\{\hat Q^{(-k)}_{t,\theta}\})$ be the nuisance
estimators computed from the data outside $I_k$ (one set of regressions for each outcome $R^{(m)}$),
with $\hat\omega^{(-k)}_t$ the corresponding weights and $\tilde Q_{t,\theta}$ the one-step conditional
expectations of \eqref{eq:remainder}. For each group $g$ (and for the whole population, $g=+$, with
$\1\{G=+\}\equiv1$ and $p_+=1$) define the remainder
\[
\mathrm{Rem}_{k,g}(\theta)=p_g^{-1}\E\Bigl[\1\{G=g\}\sum_{t=1}^T\{\hat\omega^{(-k)}_t(\theta)-\omega_t(\theta)\}\{\tilde Q_{t,\theta}-\hat Q^{(-k)}_{t,\theta}\}(H_t,A_t)\Bigr],
\]
where the expectation is over a new observation, holding the data outside $I_k$ fixed. By the argument
of Section~\ref{S-sec:proof-lemma2} applied conditionally on $G$, which is part of $H_1$,
$\E\{\1(G=g)\psi_\theta(O;\hat\eta^{(-k)})\}/p_g-V_g(\theta)=\mathrm{Rem}_{k,g}(\theta)$ exactly.

\begin{lemma}\label{S-lem:sufficient}
Suppose Assumption~\ref{ass:causal} and Condition~\ref{cond:smooth} hold, the number of folds
$K_{\rm f}$ is fixed, and for every $k$:
(i) with probability tending to one, $\hat\mu^{(-k)}_t\ge c_\mu/2$, $|\hat Q^{(-k)}_{t,\theta}|\le M'$, and
$\theta\mapsto\hat Q^{(-k)}_{t,\theta}$ is three times differentiable with derivatives bounded by $M'$;
(ii) for $\ell=0,1,2$,
$\sup_\theta\max_t\|\nabla_\theta^\ell(\hat Q^{(-k)}_{t,\theta}-\tilde Q_{t,\theta})\|=o_p(1)$, and
$\max_t\|\hat\mu_t^{(-k)}-\mu_t\|=o_p(1)$;
(iii) for every $g$, $\sup_\theta|\mathrm{Rem}_{k,g}(\theta)|=o_p(n^{-1/2})$ and
$\sup_\theta\|\nabla_\theta\mathrm{Rem}_{k,g}(\theta)\|=O_p(n^{-1/2})$.
Then Condition~\ref{cond:est}(a)--(c) holds. If $\mu$ is known and used in the weights,
$\mathrm{Rem}_{k,g}\equiv0$ and (iii) is automatic.
\end{lemma}

\begin{proof}[Sketch]
For fold $k$ with empirical measure $\mathbb{P}_{n,k}$ and each $g$,
\[
\begin{aligned}
\frac{\mathbb{P}_{n,k}\{\1(G=g)(\hat\psi_\theta-V_g(\theta))\}}{p_g}&=\frac{\mathbb{P}_{n,k}\{\1(G=g)(\psi_\theta-V_g(\theta))\}}{p_g}\\
&\quad+\frac{(\mathbb{P}_{n,k}-P_0)\{\1(G=g)(\hat\psi_\theta-\psi_\theta)\}}{p_g}+\mathrm{Rem}_{k,g}(\theta),
\end{aligned}
\]
where $\hat\psi_\theta=\psi_\theta(\cdot;\hat\eta^{(-k)})$. Since
$\hV_g(\theta)-V_g(\theta)=\Pn\{\1(G=g)(\hat\psi_\theta-V_g(\theta))\}/\Pn\1(G=g)$ and
$\Pn\1(G=g)=p_g+O_p(n^{-1/2})$, averaging over folds gives
$\hV_g(\theta)-V_g(\theta)=\Pn\varphi_{g,\theta}+o_p(n^{-1/2})$ uniformly in $\theta$ once the last two terms
are shown to be $o_p(n^{-1/2})$ uniformly.
Conditionally on the data outside $I_k$, the class $\{\hat\psi_\theta-\psi_\theta:\theta\in\Theta\}$ has a
bounded envelope and is Lipschitz in $\theta$ with a bounded constant by (i) and
Condition~\ref{cond:smooth}. Its $L_2(P_0)$ radius tends to zero. To see this, note that (ii) controls
$\hat Q_t-\tilde Q_t$, the error of each one-step regression. Since
$\tilde Q_t-Q_t=\E\{\hat V_{t+1}-V_{t+1}\mid H_t,A_t\}$ and
$\hat V_{t+1}-V_{t+1}=\sum_a\pi_{t+1}(a\mid H_{t+1})(\hat Q_{t+1}-Q_{t+1})(H_{t+1},a)$, backward induction
from $t=T$, using $\|f(H,a)\|\le c_\mu^{-1/2}\|f(H,A)\|$ by positivity, gives
$\sup_\theta\|\hat Q_{t,\theta}-Q_{t,\theta}\|=o_p(1)$ for all $t$. With (i)--(ii) this gives
$\sup_\theta\|\hat\psi_\theta-\psi_\theta\|=o_p(1)$. A maximal inequality for classes that are Lipschitz in
a finite-dimensional parameter \citep[Chapter 2.7]{van1996weak} then gives
$\sup_\theta|(\mathbb{P}_{n,k}-P_0)\{\1(G=g)(\hat\psi_\theta-\psi_\theta)\}|=o_p(n^{-1/2})$, as in
\citet{chernozhukov2018double} and \citet{kallus2020double} for a fixed $\theta$. Together with (iii) this
gives (a).

The same decomposition applied to the first and second $\theta$-derivatives, with (ii) for $\ell=1,2$,
gives (b). For the first derivative, $\sup_\theta\|\Pn\nabla_\theta\bPhi_\theta\|=O_p(n^{-1/2})$ by the
Donsker property and $\sup_\theta\|\nabla_\theta\mathrm{Rem}_{k,g}\|=O_p(n^{-1/2})$ by (iii). For the
second, $\|\nabla^\ell_\theta(\hat\omega_t-\omega_t)\|\lesssim\max_s\|\hat\mu_s-\mu_s\|$ by (i) and
Condition~\ref{cond:smooth}, so the Cauchy--Schwarz inequality and (ii) give
$\sup_\theta\|\nabla^2_\theta\mathrm{Rem}_{k,g}\|=o_p(1)$. Finally (c) follows from (i)--(ii). This is a
sketch: a complete proof would make the conditional maximal inequality and the
differentiation of $\mathrm{Rem}_{k,g}$ under the integral explicit; both are routine under (i).
\end{proof}

\section{Proof of Theorem~\ref{thm:est}}\label{S-sec:proof-thm2}

Write $\hC_j(\vth)=c_j\{\vth,\hV(\theta)\}$ and
$\Psi^\star_\vth=\Psi_{0,\vth}+\sum_{j\in\cA^\star}\lamstar_j\Psi_{j,\vth}$, so that $\Psi^\star_{\vthstar}=\phistar$.

\paragraph{Step 0 (uniform expansions).} Under Condition~\ref{cond:smooth}, $\vth\mapsto\Psi_{j,\vth}(O)$
and $\vth\mapsto\nabla_\vth\Psi_{j,\vth}(O)$ are Lipschitz with bounded constants, so these classes are
$P_0$-Donsker \citep[Example 19.7]{vandervaart1998}. Hence $\sup_\vth|\Pn\Psi_{j,\vth}|=O_p(n^{-1/2})$
and, for any $r_n\to0$,
$\sup_{\|\vth-\vth'\|\le r_n}|\Gn(\Psi_{j,\vth}-\Psi_{j,\vth'})|=o_p(1)$. A first-order Taylor expansion
of $c_j$ in $\bv$ and Condition~\ref{cond:est}(a)--(b) give
\begin{gather}
\sup_\vth|\hC_j(\vth)-C_j(\vth)-\Pn\Psi_{j,\vth}|=o_p(n^{-1/2}),\qquad \sup_\vth|\hC_j-C_j|=O_p(n^{-1/2}),\label{S-eq:s6a}\\
\sup_\vth\|\nabla\hC_j-\nabla C_j\|=O_p(n^{-1/2}),\qquad\sup_\vth\|\nabla^2\hC_j-\nabla^2C_j\|=o_p(1).\label{S-eq:s6b}
\end{gather}

\paragraph{Step 1 (consistency).} \emph{Well-separatedness:} for every $\epsilon>0$ there is
$\iota>0$ such that $\sup\{C_0(\vth):\|\vth-\vthstar\|\ge\epsilon,\ C_j(\vth)\ge-\iota\ \forall j\}<\Vstar-\iota$.
Otherwise there would be $\vth_k$ with $\|\vth_k-\vthstar\|\ge\epsilon$, $C_j(\vth_k)\ge-1/k$ and
$C_0(\vth_k)\ge\Vstar-1/k$, and a limit point would be a feasible maximiser different from
$\vthstar$. \emph{A feasible comparison point:} by linear independence there is $w$ with
$\nabla C_j(\vthstar)\T w=1$ for $j\in\cA^\star$; let $\vth_n=\vthstar+n^{-1/4}w$. Then
$C_j(\vth_n)=n^{-1/4}+O(n^{-1/2})$ for $j\in\cA^\star$ and $C_j(\vth_n)\to C_j(\vthstar)>0$ otherwise, so by
\eqref{S-eq:s6a} $\vth_n$ is feasible for the empirical problem with probability tending to one. Hence
$\hC_0(\hvth)\ge\hC_0(\vth_n)=\Vstar+o_p(1)$, so $C_0(\hvth)\ge\Vstar-o_p(1)$; also
$C_j(\hvth)\ge-O_p(n^{-1/2})$. By well-separatedness, $\hvth\to\vthstar$ in probability.

\paragraph{Step 2 (KKT, active set and multipliers).} With probability tending to one, $\hvth$ is
interior, $\hC_j(\hvth)\ge C_j(\vthstar)/2>0$ for $j\notin\cA^\star$, and the gradients
$\nabla\hC_j(\hvth)$, $j\in\cA^\star$, are linearly independent (they converge to
$\nabla C_j(\vthstar)$). The first-order necessary conditions therefore hold at $\hvth$ with unique
multipliers: $\hlam_j=0$ for $j\notin\cA^\star$ and
$\hlam_{\cA^\star}=-(\hat{\mathbf{D}}\T\hat{\mathbf{D}})^{-1}\hat{\mathbf{D}}\T\nabla\hC_0(\hvth)$, where $\hat{\mathbf{D}}$ has columns $\nabla\hC_j(\hvth)$,
$j\in\cA^\star$. By \eqref{S-eq:s6b} and consistency, $\hlam_{\cA^\star}\to\lamstar_{\cA^\star}$, so $\hlam_j>0$ and, by
complementary slackness, $\hC_j(\hvth)=0$ for all $j\in\cA^\star$, with probability tending to one. The
estimated active set of Algorithm~\ref{alg:main} is then $\cA^\star$ (for the fixed tolerance $\tau$ of Theorem~\ref{thm:est}(i)), and the non-negative least squares
step returns $\hlam$.

\paragraph{Step 3 ($\sqrt n$-rate).} Let
$\hat F(\vth,\lambda)=(\nabla\hC_0+\sum_{j\in\cA^\star}\lambda_j\nabla\hC_j,(\hC_j)_{j\in\cA^\star})$ and $F$ its
population analogue. Then $\hat F(\hvth,\hlam_{\cA^\star})=0=F(\vthstar,\lamstar_{\cA^\star})$, so
\[
0=F(\hvth,\hlam_{\cA^\star})-F(\vthstar,\lamstar_{\cA^\star})+(\hat F-F)(\hvth,\hlam_{\cA^\star}).
\]
Write $x=(\vth,\lambda_{\cA^\star})$, $\hat x=(\hvth,\hlam_{\cA^\star})$ and $x^\star=(\vthstar,\lamstar_{\cA^\star})$.
The last term is $O_p(r_n)$, where $r_n$ bounds the uniform gradient error in
Condition~\ref{cond:est}(b); by \eqref{S-eq:s6a}--\eqref{S-eq:s6b}, $r_n=n^{-1/2}$ under that condition. By
the integral form of the mean-value theorem,
$F(\hat x)-F(x^\star)=\{\int_0^1DF(x^\star+u(\hat x-x^\star))\,du\}(\hat x-x^\star)=\{\Kstar+o_p(1)\}(\hat x-x^\star)$,
because $DF$ is continuous and $\hat x\to x^\star$. Since $\Kstar$ is nonsingular,
$\|\hvth-\vthstar\|+\|\hlam-\lamstar\|=O_p(r_n)$, which is $O_p(n^{-1/2})$ under Condition~\ref{cond:est}(b)
and $o_p(n^{-1/4})$ if the gradient error is only $o_p(n^{-1/4})$. Note that only the population Jacobian
enters here; the Hessian part of Condition~\ref{cond:est}(b) is not used.

\paragraph{Step 4 (asymptotic linearity).} By Step 2, with probability tending to one,
\[
\hV^\star=\hC_0(\hvth)=\hC_0(\hvth)+\sum_{j\in\cA^\star}\lamstar_j\hC_j(\hvth)=\hat L(\hvth),\qquad
\hat L(\vth):=\hC_0(\vth)+\sum_{j\in\cA^\star}\lamstar_j\hC_j(\vth),
\]
where we stress that the \emph{true} multipliers are used. With $L(\vth)=C_0(\vth)+\sum_{j\in\cA^\star}\lamstar_jC_j(\vth)$,
\[
\hV^\star-\Vstar=\{L(\hvth)-L(\vthstar)\}+(\hat L-L)(\hvth),
\]
because $L(\vthstar)=C_0(\vthstar)=\Vstar$. Since $\nabla L(\vthstar)=0$ (stationarity),
$L(\hvth)-L(\vthstar)=O(\|\hvth-\vthstar\|^2)$, which is $O_p(n^{-1})$ under Condition~\ref{cond:est}(b)
and $o_p(n^{-1/2})$ in any case by Step 3.
By \eqref{S-eq:s6a}, consistency and stochastic equicontinuity,
$(\hat L-L)(\hvth)=\Pn\Psi^\star_{\hvth}+o_p(n^{-1/2})=\Pn\Psi^\star_{\vthstar}+o_p(n^{-1/2})$, using
$P_0\Psi^\star_\vth=0$. Hence $\sqrt n(\hV^\star-\Vstar)=\Gn\phistar+o_p(1)$. Note that only
$\|\hvth-\vthstar\|=o_p(n^{-1/4})$ was used. An asymptotically linear estimator whose influence function
is the EIF is regular and efficient \citep[Lemma 25.23]{vandervaart1998}, which proves (ii); (i) is
Steps 1--3.

\paragraph{Step 5 (variance estimator).} With $\hat\varphi^\star_i$ as in Algorithm~\ref{alg:main},
$\Pn(\hat\varphi^\star-\phistar)^2\le2\Pn(\hat\varphi^\star-\tilde\varphi)^2+2\Pn(\tilde\varphi-\phistar)^2$,
where $\tilde\varphi=\sum_j\hlam_j\nabla_{\bv}c_j\{\hvth,\hV(\hth)\}\T\bPhi_{\hth}$. The first term is
$o_p(1)$ by Condition~\ref{cond:est}(c) and boundedness of $\nabla c_j$ and $\hlam$; the second by
consistency of $(\hvth,\hlam,\hV)$, the Lipschitz property in $\vth$ and a uniform law of large
numbers. Hence $\hat\sigma^2\to\E(\phistar{}^2)$, giving (iii).

\paragraph{Step 6 (price of fairness).} The unconstrained problem has a unique global maximiser in the
interior with negative definite Hessian (Corollary~\ref{cor:other}(a)), and $\hat\theta_{\rm unc}$ is a
global maximiser of $\hV_+$; so Steps 1--4 with $J=0$ give
$\sqrt n(\hC_0(\hat\theta_{\rm unc})-V^\star_{\rm unc})=\Gn\varphi_{+,\theta_{\rm unc}}+o_p(1)$. Subtracting the
expansion of Step 4 proves the asymptotic linearity in (iv), and the argument of Step 5 applied to
$\hat\varphi_{+,\hat\theta_{\rm unc}}-\hat\varphi^\star$ gives consistency of the variance estimator.

\paragraph{Step 7 (optimiser and multipliers).} Under the additional condition,
\[
(\hat F-F)(\vthstar,\lamstar_{\cA^\star})=\Pn\bigl(\nabla\Psi^\star_{\vthstar},(\Psi_{j,\vthstar})_{j\in\cA^\star}\bigr)+o_p(n^{-1/2})
\]
and, by stochastic equicontinuity and Step 3,
$(\hat F-F)(\hvth,\hlam_{\cA^\star})-(\hat F-F)(\vthstar,\lamstar_{\cA^\star})=o_p(n^{-1/2})$. The display in Step 3
then gives
\[
(\hvth,\hlam_{\cA^\star})-(\vthstar,\lamstar_{\cA^\star})=-(\Kstar)^{-1}\Pn\bigl(\nabla\Psi^\star_{\vthstar},(\Psi_{j,\vthstar})_{j\in\cA^\star}\bigr)+o_p(n^{-1/2}),
\]
which is \eqref{eq:if-theta-lambda}. The Corollary~\ref{cor:other}(b) functionals follow by the
delta method. The variance estimators in (v) are consistent by the argument of Step 5, using
$\sup_\vth\|\nabla^2\hC_j-\nabla^2C_j\|=o_p(1)$, a difference quotient with step $h_n\to0$ chosen so that
the differencing error of $\nabla\hC_j$ is $o_p(1)$, and the added condition on $\nabla_\theta\hat\bPhi_\theta$.

\paragraph{Step 8 (part (vi)).} With $\bar\mu=\mu$, the remainder \eqref{eq:remainder} vanishes, so
$P_0\psi_\theta(\cdot;\mu,\bar Q)=V_+(\theta)$ for every $\theta$, and similarly for the group values. Hence
every component of $\bar\bPhi_\theta$ has mean zero, $P_0\bar\Psi_{j,\vth}=0$ for all $\vth$, and
$\vth\mapsto\bar\Psi_{j,\vth}$ is Lipschitz. These are the only properties of $\bPhi$ used in Steps 0--6
and in the proof of Proposition~\ref{prop:cert}, which therefore hold with $\bar\bPhi$ in place of
$\bPhi$; the Lipschitz property follows from the assumption in (vi) that each $\bar Q_{t,\theta}$ is
bounded and Lipschitz in $\theta$ uniformly in $(h,a)$, together with Condition~\ref{cond:smooth}.
(The efficiency statement in (ii), which relies on $\phistar$ being the EIF, is not carried
over.) Write $\Delta_t=\bar Q_t-Q_t$ and $b_t=\omega_t\Delta_t(H_t,A_t)
=\omega_{t-1}\pi_{\theta,t}(A_t\mid H_t)\Delta_t(H_t,A_t)/\mu_t(A_t\mid H_t)$. From
$\psi=\omega_TR-\sum_t\{\omega_tQ_t(H_t,A_t)-\omega_{t-1}V_t(H_t)\}$,
\[
\psi_\theta(\cdot;\mu,\bar Q)-\psi_\theta(\cdot;\mu,Q)=-\sum_{t=1}^T\bigl\{b_t-\E(b_t\mid H_t)\bigr\},
\]
an element of the closed span of the logging-policy scores. The same holds for the group versions, and
for $s_g$ the sample-mean influence function differs from the known-$\mu$ EIF by such an element
(Section~\ref{S-sec:proof-thm1}, Step 4). Hence $\bar\varphi^\star-\phistar$ is orthogonal to the tangent space
of the model with $\mu$ known, so $\bar\varphi^\star$ is a gradient there. An asymptotically linear
estimator whose influence function is a gradient is regular \citep[Chapter 25]{vandervaart1998}, but it
is efficient only if that gradient is the canonical one, that is, the known-$\mu$ efficient influence
function of Remark~\ref{rem:knownmu}. This holds if $\bar Q_{\cdot,\thstar}=Q_{\cdot,\thstar}$ and no $s_g$
enters the criterion, and fails under generic misspecification. The condition is sufficient but not
necessary: $\bar Q$ may differ from $Q$ in directions that leave $\psi$ unchanged (for instance
$\bar Q_t=Q_t+c_t(H_t)\mu_t/\pi_{\thstar,t}$, for which $b_t=\E(b_t\mid H_t)$), or in groups whose
multiplier is zero.
\qed

\section{Proof of Proposition~\ref{prop:cert}}\label{S-sec:proof-prop}

(a) By Step 2 of Section~\ref{S-sec:proof-thm2}, $\hC_1(\hvth)=0$ with probability tending to one, so by
\eqref{S-eq:s6a} and stochastic equicontinuity
$C_1(\hvth)=-(\hC_1-C_1)(\hvth)=-\Pn\Psi_{1,\vthstar}+o_p(n^{-1/2})$. By the central limit theorem,
$\sqrt nC_1(\hvth)\rightsquigarrow N(0,\sigma_1^2)$, a continuous law symmetric about zero, so
$\Pr\{C_1(\hvth)\ge0\}\to1/2$. Moreover, since $C_0=L-\lamstar C_1$ and $L(\hvth)-\Vstar=O(\|\hvth-\vthstar\|^2)=o_p(n^{-1/2})$,
$C_0(\hvth)=\Vstar+\lamstar\Pn\Psi_{1,\vthstar}+o_p(n^{-1/2})$. If other constraints are present, they are slack
at $\vthstar$ and remain inactive at $\hvth$ with probability tending to one (Step 2), so they play no role.

(b) Let $k_n(\vth)=z_{1-\alpha}\hat\sigma_1(\vth)n^{-1/2}$. Since $\sup_\vth\hat\sigma_1=O_p(1)$ by assumption,
$\sup_\vth|k_n|=O_p(n^{-1/2})$ globally, which is all that Step 1 (consistency) requires. The argument of
Step 5 of Section~\ref{S-sec:proof-thm2} gives $\sup_\vth|\hat\sigma^2_1(\vth)-\sigma_1^2(\vth)|=o_p(1)$ with
$\sigma_1^2(\vth)=\Var(\Psi_{1,\vth})$ continuous; as $\sigma_1=\sigma_1(\vthstar)>0$, $\hat\sigma_1$ is bounded away
from zero near $\vthstar$ with probability tending to one, and by the stated smoothness of $\hat\sigma_1^2$,
$\sup(\|\nabla k_n\|+\|\nabla^2k_n\|)=O_p(n^{-1/2})$ over a neighbourhood of $\vthstar$,
so \eqref{S-eq:s6a}--\eqref{S-eq:s6b} hold for $\hC_1-k_n$, with $\Pn\Psi_{1,\vth}$ replaced by
$\Pn\Psi_{1,\vth}-k_n(\vth)$ in the first display. Steps 1--3 of Section~\ref{S-sec:proof-thm2} then apply
verbatim to the tightened problem: $\hvth_\alpha\to\vthstar$ at rate $n^{-1/2}$ and
$\hC_1(\hvth_\alpha)=k_n(\hvth_\alpha)$ with probability tending to one. Hence
\[
C_1(\hvth_\alpha)=k_n(\hvth_\alpha)-(\hC_1-C_1)(\hvth_\alpha)=n^{-1/2}\{z_{1-\alpha}\sigma_1-\Gn\Psi_{1,\vthstar}\}+o_p(n^{-1/2}),
\]
using $\hat\sigma_1(\hvth_\alpha)\to\sigma_1$, which follows from the uniform consistency of $\hat\sigma_1^2$, the
continuity of $\sigma_1^2(\cdot)$ and $\hvth_\alpha\to\vthstar$. Therefore
$\Pr\{C_1(\hvth_\alpha)\ge0\}\to\Pr(Z\le z_{1-\alpha})=1-\alpha$ for $Z\sim N(0,1)$, and
$C_0(\hvth_\alpha)=L(\hvth_\alpha)-\lamstar C_1(\hvth_\alpha)=\Vstar-\lamstar n^{-1/2}\{z_{1-\alpha}\sigma_1-\Gn\Psi_{1,\vthstar}\}+o_p(n^{-1/2})$.
\qed

\section{Exact numerical verification}\label{S-sec:exact}

\input{tables/tex_exact_supp.tex}

\paragraph{General criteria.} A second check uses a different discrete law with three
groups ($p_g=0.45,0.35,0.20$), binary $X_1,A_1,X_2,A_2,R$ (96 cells) and the three-parameter class
$\pi_{\theta,1}(1\mid H_1)=\expit\{4(X_1-\theta_1)\}$,
$\pi_{\theta,2}(1\mid H_2)=\expit[4\{(X_1+X_2)/2-\theta_{2+X_1}\}]$. The functional vector contains the
group values $V_g$, the intervention rates $D_g$ (auxiliary outcome $(A_1+A_2)/2$), the proportions $p_g$
and, for criterion II, the status-quo means $s_g$. Criterion I maximises the inequality-averse welfare
$(\sum_gp_gV_g^{-1})^{-1}$ subject to $V_1/V_0\ge0.7935$ (binding, $\lamstar=0.377$),
$D_0-D_2\le0.165$ (binding, $\lamstar=0.193$) and a slack budget $\sum_gp_gD_g\le b$. Criterion II
maximises $\min_g(V_g-s_g)$ through the epigraph variable $u$; two of the three constraints are active
($\lamstar=0.879, 0.121$, summing to one). In both cases a dense grid search confirms global
optimality, linear independence holds and the reduced Hessian is negative definite. Perturbed problems are
solved by Newton's method on the KKT system with complex-step gradients, and derivatives are
Richardson-extrapolated central differences. For criterion II in the model with $\mu$ known, the random directions
are projected onto the orthocomplement of the action scores and the known-$\mu$ influence function of
$s_g$ (Lemma~\ref{lem:eif-fixed}) is used. Table~\ref{S-tab:exact-general} reports the results. For
criterion I, the wrong candidates give errors of $0.10$ (no multiplier term),
$0.018$ (adding a slack constraint's term), $0.40$ ($V_g(\thstar)$ with $\thstar$ treated as fixed) and
$0.43$ (multiplier influence function that ignores the curvature of $c_j$).

\input{tables/tex_exact_general.tex}

\section{Simulation: details and additional results}\label{S-sec:sim-details}

\input{tables/tex_sim_supp.tex}

\FloatBarrier
\section{Case study: details}\label{S-sec:case-details}

\input{tables/tex_case_supp.tex}

\FloatBarrier
\section{Additional remarks}\label{S-sec:remarks}

\paragraph{Comparison of efficiency bounds.} For the running example \eqref{eq:eif-example},
\[
\Var(\phistar)=\Var(\varphi_{+,\thstar})+2\lamstar\Cov(\varphi_{+,\thstar},\varphi_{A,\thstar}-\varphi_{B,\thstar})
+\lamstar{}^2\Var(\varphi_{A,\thstar}-\varphi_{B,\thstar}).
\]
Because $\E(\varphi_{g,\theta}\mid G)=0$ and the groups are disjoint, with
$\sigma_g^2=\Var(\psi_{\thstar}\mid G=g)$ this equals
\[
\Var(\varphi_{+,\thstar})+2\lamstar(\sigma_A^2-\sigma_B^2)+\lamstar{}^2\bigl(\sigma_A^2/p_A+\sigma_B^2/p_B\bigr).
\]
The additional terms reflect the estimation of the binding constraint. If $\sigma_B^2>\sigma_A^2$ and
$\lamstar<2(\sigma_B^2-\sigma_A^2)/(\sigma_A^2/p_A+\sigma_B^2/p_B)$, the bound is smaller than
$\Var(\varphi_{+,\thstar})$; otherwise it is larger. Since the bounds for $\Vstar(\delta)$ and
$V^\star_{\rm unc}$ are moreover evaluated at different policies, there is no general ordering between
the constrained and unconstrained bounds.

\paragraph{Equality constraints.} Equality constraints, such as $V_A=V_B$ (criterion (c) of
Table~\ref{tab:criteria} with $\delta=0$), can be handled by the same arguments. In
Theorem~\ref{thm:eif} their multipliers have unrestricted sign and no complementarity condition is
needed; in the proof of Theorem~\ref{thm:est} the feasible comparison point is obtained from the
implicit function theorem, and the multipliers are estimated by unconstrained least squares.

\paragraph{Numerical derivatives.} The theory treats exact derivatives. The central differences used in
Algorithm~\ref{alg:main} introduce errors of order $h^2$, negligible for the step $h=10^{-4}$ used
here, and the difference-quotient estimate of $\Kstar$ with a fixed step is a numerical approximation
to the consistent estimator of Theorem~\ref{thm:est}(v), which requires a step $h_n\to0$.

\FloatBarrier

\end{document}

%% file: macros.tex
\newcommand{\E}{\mathbb{E}}
\newcommand{\R}{\mathbb{R}}
\renewcommand{\Pr}{\mathbb{P}}
\newcommand{\Pn}{\mathbb{P}_n}
\newcommand{\Gn}{\mathbb{G}_n}
\newcommand{\1}{\mathbf{1}}
\newcommand{\bV}{\mathbf{V}}
\newcommand{\bv}{\mathbf{v}}
\newcommand{\bPhi}{\boldsymbol{\Phi}}
\newcommand{\vth}{\vartheta}
\newcommand{\cA}{\mathcal{A}}
\newcommand{\cG}{\mathcal{G}}

\newcommand{\cM}{\mathcal{M}}

\newcommand{\cU}{\mathcal{U}}
\newcommand{\Var}{\operatorname{Var}}
\newcommand{\Cov}{\operatorname{Cov}}
\newcommand{\PoF}{\operatorname{PoF}}
\newcommand{\expit}{\operatorname{expit}}

\newcommand{\Vstar}{V^{\star}}
\newcommand{\thstar}{\theta^{\star}}
\newcommand{\vthstar}{\vartheta^{\star}}
\newcommand{\lamstar}{\lambda^{\star}}
\newcommand{\phistar}{\varphi^{\star}}
\newcommand{\Kstar}{\mathbf{K}^{\star}}
\newcommand{\hV}{\widehat{V}}
\newcommand{\hC}{\widehat{C}}
\newcommand{\hth}{\widehat{\theta}}
\newcommand{\hvth}{\widehat{\vartheta}}
\newcommand{\hlam}{\widehat{\lambda}}
\newcommand{\T}{^{\top}}
\newcommand{\Aset}{\mathbb{A}}

%% file: tables/tex_exact_lambda.tex
$\lamstar=0.328$

%% file: tables/tex_exact.tex
\begin{tabular}{llcc}
\toprule
Functional & Influence function used & Max.\ error & Max.\ $|$derivative$|$\\
\midrule
$\Vstar$ & Theorem 1, $\varphi_++\lamstar(\varphi_A-\varphi_B)$ & $2.6\times10^{-11}$ & 0.124\\
$\Vstar$ & constraint ignored, $\varphi_+$ & 0.056 & 0.124\\
$\thstar_1$ & Theorem 1(ii) & $2.1\times10^{-9}$ & 3.129\\
$\thstar_2$ & Theorem 1(ii) & $1.6\times10^{-9}$ & 2.675\\
$\lamstar$ & Theorem 1(ii) & $6.7\times10^{-10}$ & 1.524\\
$V_A(\thstar)$ & Corollary 1(b) & $2.0\times10^{-11}$ & 0.123\\
$V_A(\thstar)$ & $\thstar$ treated as fixed, $\varphi_{A}$ & 0.047 & 0.123\\
$\PoF$ & Corollary 1(a) & $2.4\times10^{-11}$ & 0.059\\
\bottomrule
\end{tabular}

%% file: tables/tex_sim_main.tex
\begin{table}[t]
\centering
\caption{Monte Carlo results for the constrained optimal value and the price of fairness
(500 replications for each $n$, excluding the few in which the empirical problem was infeasible;
see the text). Bias, standard deviation (SD) across replications and mean
estimated standard error (SE) are multiplied by 100; Cov is the coverage (\%) of nominal 95\% Wald
intervals. True values: Table~\ref{S-tab:truth}.}
\label{tab:main}
\footnotesize
\setlength{\tabcolsep}{4pt}
\begin{tabular}{@{}cc ccc c cc c ccc c@{}}
\toprule
 & & \multicolumn{4}{c}{$\Vstar(\delta)$, doubly robust} & \multicolumn{2}{c}{$\Vstar(\delta)$, IPW} & & \multicolumn{4}{c}{$\PoF(\delta)$, doubly robust}\\
\cmidrule(lr){3-6}\cmidrule(lr){7-8}\cmidrule(lr){10-13}
$n$ & $\delta$ & Bias & SD & SE & Cov & SD & Cov & & Bias & SD & SE & Cov\\
\midrule
1000 & 0.5 & 0.6 & 5.5 & 5.9 & 97.2 & 6.9 & 95.1 &  & 0.3 & 4.5 & 4.9 & 90.0\\
 & 0.6 & 0.4 & 4.9 & 5.2 & 96.2 & 6.4 & 97.4 &  & 0.5 & 3.1 & 3.4 & 86.2\\
 & 0.7 & 0.3 & 4.7 & 4.9 & 96.4 & 6.2 & 97.6 &  & 0.6 & 1.9 & 1.9 & 77.0\\
 & 0.8 & 0.4 & 4.7 & 4.9 & 96.4 & 6.2 & 96.4 &  & 0.5 & 0.9 & 0.9 & 56.8\\
 & 1.0 & 0.8 & 4.8 & 5.0 & 96.4 & 6.7 & 95.2 &  & 0.0 & 0.1 & 0.0 & --\\
\addlinespace
2000 & 0.5 & -0.1 & 3.9 & 4.1 & 95.8 & 4.5 & 95.6 &  & 0.3 & 3.4 & 3.5 & 91.0\\
 & 0.6 & -0.1 & 3.5 & 3.6 & 95.8 & 4.0 & 97.4 &  & 0.4 & 2.4 & 2.4 & 90.2\\
 & 0.7 & -0.2 & 3.2 & 3.4 & 96.6 & 3.8 & 97.6 &  & 0.4 & 1.4 & 1.3 & 84.0\\
 & 0.8 & -0.1 & 3.2 & 3.4 & 96.6 & 4.0 & 97.8 &  & 0.3 & 0.6 & 0.5 & 62.0\\
 & 1.0 & 0.2 & 3.3 & 3.5 & 97.0 & 4.3 & 97.4 &  & 0.0 & 0.0 & 0.0 & --\\
\addlinespace
4000 & 0.5 & -0.3 & 2.9 & 2.9 & 94.8 & 3.4 & 94.8 &  & 0.2 & 2.5 & 2.5 & 93.6\\
 & 0.6 & -0.2 & 2.5 & 2.5 & 95.6 & 3.1 & 95.4 &  & 0.2 & 1.7 & 1.7 & 92.2\\
 & 0.7 & -0.2 & 2.3 & 2.4 & 96.6 & 3.0 & 94.6 &  & 0.2 & 0.9 & 0.9 & 87.8\\
 & 0.8 & -0.2 & 2.4 & 2.4 & 96.0 & 3.0 & 94.0 &  & 0.1 & 0.3 & 0.3 & 65.2\\
 & 1.0 & 0.0 & 2.4 & 2.5 & 95.8 & 3.1 & 93.8 &  & 0.0 & 0.0 & 0.0 & --\\
\bottomrule
\end{tabular}
\end{table}

\begin{table}[t]
\centering
\caption{Inference beyond the optimal value, doubly robust estimator, $n=4000$. Coverage (\%) of
95\% intervals for $\Vstar$ with the standard error from Theorem~\ref{thm:eif} and from the influence
function that ignores the constraint (naive); bias, ratio of Monte Carlo SD to mean SE, and coverage for $\lamstar$;
coverage for the value delivered to group $A$, $V_A(\thstar)$, using Corollary~\ref{cor:other} or
treating $\hat\theta$ as fixed; percentage of replications in which the estimated policy satisfies the
\emph{population} constraint, for the plug-in and the certified ($\alpha=0.05$) policies (a
replication in which the tightened problem of Proposition~\ref{prop:cert}(b) has no solution in $\Theta$
counts as not feasible); and mean true value loss $\Vstar-V(\hat\theta)$ ($\times100$) over the
replications in which the policy exists.}
\label{tab:inf}
\footnotesize
\setlength{\tabcolsep}{3pt}
\begin{tabular}{@{}c cc c cccc c cc c cc cc@{}}
\toprule
 & \multicolumn{2}{c}{Cov.\ of $\Vstar$ interval} & & \multicolumn{4}{c}{Multiplier $\lamstar$} & & \multicolumn{2}{c}{Cov.\ $V_A(\thstar)$} & & \multicolumn{2}{c}{$\Pr\{C_1(\hat\theta)\ge0\}$} & \multicolumn{2}{c}{Value loss}\\
\cmidrule(lr){2-3}\cmidrule(lr){5-8}\cmidrule(lr){10-11}\cmidrule(lr){13-14}\cmidrule(lr){15-16}
$\delta$ & Thm 1 & naive & & True & Bias & SD/SE & Cov & & Cor.~1 & naive & & plug-in & certified & plug-in & certified\\
\midrule
0.5 & 94.8 & 91.8 &  & 0.412 & 0.001 & 0.99 & 93.8 &  & 94.6 & 97.4 &  & 45.8 & 91.8 & 0.26 & 4.33\\
0.6 & 95.6 & 94.8 &  & 0.291 & 0.001 & 1.01 & 95.0 &  & 95.0 & 98.8 &  & 45.8 & 93.8 & 0.21 & 3.02\\
0.7 & 96.6 & 97.0 &  & 0.165 & 0.001 & 1.01 & 94.4 &  & 96.2 & 98.8 &  & 47.6 & 94.2 & 0.20 & 1.85\\
0.8 & 96.0 & 96.2 &  & 0.035 & 0.015 & 1.04 & 65.6 &  & 95.4 & 97.0 &  & 58.2 & 96.4 & 0.23 & 0.84\\
\bottomrule
\end{tabular}
\end{table}

%% file: tables/tex_sim_text.tex
Table~\ref{tab:main} summarises estimation of $\Vstar(\delta)$ and $\PoF(\delta)$, and
Table~\ref{tab:inf} reports the other inferential targets at $n=4000$; Section~\ref{S-sec:sim-details}
of the Supplement gives the corresponding results for $n=1000$ and $2000$. In a few replications at
$\delta=0.5$ the empirical problem had no feasible solution (DR: 2 of 500 at $n=1000$ and at
$n=2000$; IPW: at most 12 of 500), and these replications are excluded.

\emph{Constrained value.} The doubly robust estimator of $\Vstar(\delta)$ is essentially unbiased
(at most 0.008 in absolute value, against standard deviations of 0.023--0.055), and its Wald
intervals cover at 94.8--97.2\%, including the near-boundary case $\delta=0.8$ and the slack case
$\delta=1$; as Remark~\ref{rem:nonregular} anticipates, the interval for $\Vstar$ is not affected near
the boundary. The intervals are slightly conservative at $n\le2000$ because the
cross-fitted standard error overstates the Monte Carlo standard deviation by a few per cent; the same
happens for a fixed policy (Table~\ref{S-tab:gapfixed}). The IPW estimator is also valid, but its
standard deviation is 15--40\% larger, a variance ratio of about 1.3--2. This is the efficiency gain
expected from the doubly robust form, although the regressions are misspecified.

\emph{Which influence function?} Table~\ref{tab:inf} compares intervals for $\Vstar$ built from the
influence function of Theorem~\ref{thm:eif} with naive intervals that ignore the constraint. With
Theorem~\ref{thm:eif}, coverage is 94.8--96.6\%. The naive intervals undercover when the multiplier is
large (91.8\% at $\delta=0.5$); the difference shrinks as $\lamstar$ decreases, as
\eqref{eq:eif-example} implies. The exact check in Table~\ref{tab:exact} confirms, without sampling
noise, that the naive influence function is incorrect.

\emph{Price of fairness, multiplier and group values.} Coverage of the $\PoF$ intervals improves
with $n$ when the constraint binds firmly: from 90.0\% to 93.6\% at $\delta=0.5$ and from 86.2\% to
92.2\% at $\delta=0.6$ as $n$ grows from 1000 to 4000. At $\delta=0.7$, where $\PoF=0.011$ is about one
standard deviation from zero, coverage is lower (77.0\% at $n=1000$, 87.8\% at $n=4000$), and the
positive bias of the plug-in estimator contributes. It remains poor at $\delta=0.8$, where
$\PoF=0.0005$ and $\lamstar=0.035$ are within a fraction of a standard error of zero, and the
constraint is active in only about two thirds of the samples. This is the nonregular regime of
Remark~\ref{rem:nonregular}(ii), where Wald intervals for the price of fairness and the multiplier
should not be trusted. For $\delta\le0.7$ the multiplier is estimated essentially without bias and its
KKT-based intervals cover at 93.8--95.0\%; at $\delta=0.8$ the bias (0.015) is not small relative to
$\lamstar=0.035$. For the value delivered to group $A$, the Corollary~\ref{cor:other}
intervals cover at 94.6--96.2\%, whereas treating $\hat\theta$ as fixed gives intervals that are too
wide here (97.0--98.8\%). In the case study of Section~\ref{sec:case}, the naive standard errors are
too small for some criteria and too large for others.

\emph{Feasibility and certification.} Consistent with Proposition~\ref{prop:cert}(a), when the
constraint binds firmly ($\delta\le0.7$) the plug-in fair policy satisfies the \emph{population}
constraint in 45.8--47.6\% of replications at $n=4000$ (40.4--52.2\% at $n\le2000$), close to one
half; the small shortfall is a second-order effect. At $\delta=0.8$, where the estimated constraint is
often inactive, the rate is higher (58--68\%). The mean true value of the plug-in policy is below
$\Vstar$ by a second-order amount that decreases with $n$ (0.005--0.009 at $n=1000$, 0.002--0.003 at
$n=4000$). The certified policy of Proposition~\ref{prop:cert}(b) is feasible in 91.8--96.4\% of
replications at $n=4000$. At smaller $n$ and $\delta=0.5$ the tightened empirical problem often has
no solution in $\Theta$, so that no policy can be certified (146 of 498 replications at $n=1000$ and
47 at $n=2000$). Counting these as failures, the rates are 62.4--95.4\% at $n=1000$ and 84.7--97.4\%
at $n=2000$; among the replications in which a certified policy exists they are 88.4--95.4\% and
93.6--97.4\% (Section~\ref{S-sec:sim-details}). The value loss of the certified policy at $n=4000$ is $0.043$,
$0.030$, $0.019$ and $0.008$ for $\delta=0.5,\ldots,0.8$, decreasing with $\lamstar$ and with $n$, as
Proposition~\ref{prop:cert}(b) predicts.

\emph{Nuisance estimation and misspecification.} Table~\ref{S-tab:robust} in the Supplement
reports the three additional scenarios at $n=2000$. With logistic propensities and series
regressions (S2), the results are almost identical to those with the logged propensities. With a
constant outcome model and correct propensities (S3), or with series regressions and propensities
that omit $X_t$ and $A_{t-1}$ (S4), the doubly robust estimator of $\Vstar$ remains nearly unbiased
(absolute bias at most 0.009, against standard deviations of 0.041--0.049) and its intervals cover at
93.7--98.3\%, at the cost of a 5--34\% larger standard deviation. For S3 this reflects the double
robustness in \eqref{eq:remainder} through the correct propensity model; for S4 it reflects that the
series regressions are close to correct. The over-coverage reflects conservative standard errors. Coverage for the
price of fairness follows the same pattern as in S1.

%% file: tables/tex_case.tex
\emph{All data in this section are synthetic.} The aim is to show how the methods would be used on an
adaptive-tutoring log; the generating model is given in Section~\ref{S-sec:case-details} of the
Supplement. Because the data are synthetic, we can also report the true value of every estimated
policy, which a real analysis could not.

\paragraph{Data.} We simulated logs for $n=5000$ students over $T=4$ sessions. Of these students,
35\% belong to group $A$ (English learners) and start with lower mastery and lower
engagement. The platform randomises between ``advance'' and ``review'' with logged probabilities in
$[0.1,0.9]$ that depend on mastery and engagement. The benefit of advancing grows with current and
previous mastery and with engagement, and is smaller in group $A$; end-of-course mastery $R$ is the
outcome. Policies are soft-threshold rules \eqref{eq:softthreshold} on recent average mastery, with
one threshold for sessions 1--2 and one for sessions 3--4. They use neither $G$ nor engagement, so
they cannot target group $A$ directly. We used the doubly robust estimator with the logged
propensities, $K_{\rm f}=5$ folds and series regressions, and Algorithm~\ref{alg:main}.

\begin{figure}[t]
\centering
\includegraphics[width=\textwidth]{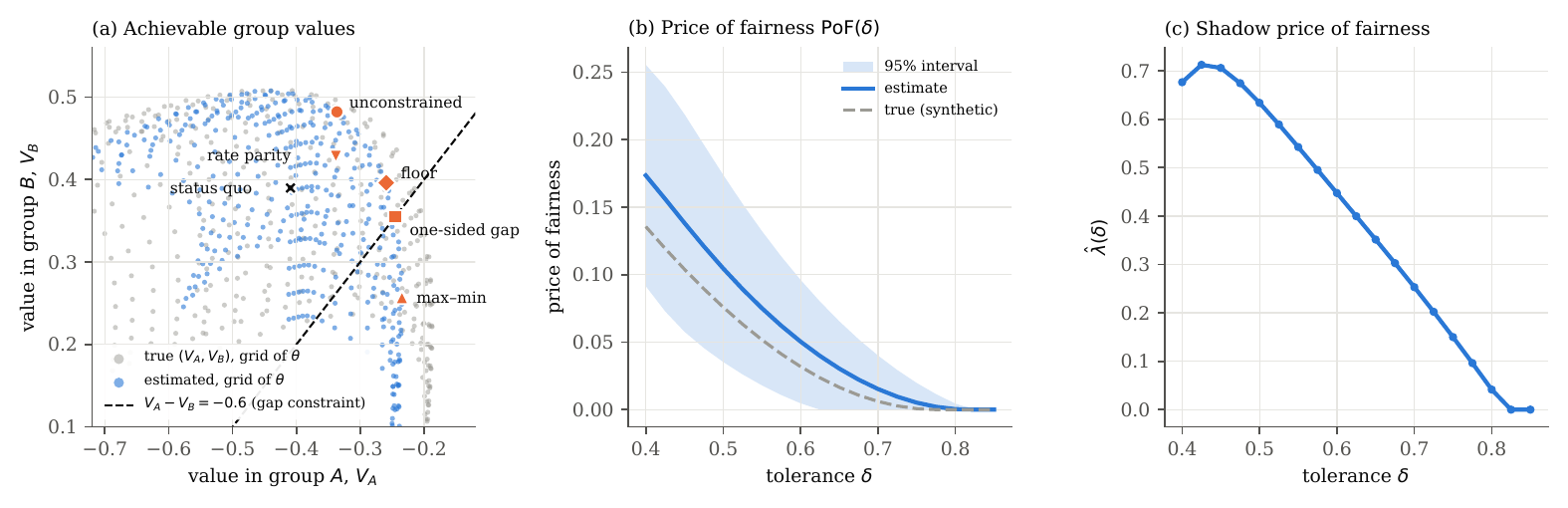}
\caption{Synthetic case study. (a) Estimated (blue) and true (grey) group values $(V_A,V_B)$ over a
$21\times21$ grid of policies; the dashed line is the boundary of the gap constraint at
$\delta=0.6$; markers show the estimated optima under the criteria of Table~\ref{tab:case} and the
status quo. (b) Estimated price of fairness with pointwise 95\% intervals (truncated at zero) and the
true curve. (c) Estimated multiplier (shadow price) $\hat\lambda(\delta)=-\partial\PoF/\partial\delta$.}
\label{fig:case}
\end{figure}

\paragraph{Unconstrained optimum and the gap.} The unconstrained optimal policy has estimated value
$0.194$ (SE $0.022$), but it leaves group $A$ far behind: $\hat V_A-\hat V_B=-0.819$
(SE $0.050$). The one-sided level-0.05 test of Section~\ref{sec:inference} therefore rejects
the hypothesis that the unconstrained optimum satisfies the gap constraint for every
$\delta<0.74$. Figure~\ref{fig:case}(a) shows the estimated and true achievable sets of
group values over the policy class. The trade-off is clear: thresholds that help group $A$, by
advancing fewer weak students, cost group $B$ more than they gain for $A$.

\paragraph{Price of fairness.} Figure~\ref{fig:case}(b) shows $\widehat\PoF(\delta)$ with pointwise
95\% intervals over $\delta\in[0.4,0.85]$, together with the true curve, and panel (c) shows the
estimated shadow price $\hat\lambda(\delta)$. At $\delta=0.6$ the estimated constrained value is
$0.143$ (SE $0.028$). The naive standard error that ignores the constraint would be
$0.023$. The price of fairness is $0.050$ (SE $0.023$; true value $0.032$), and
$\hat\lambda=0.45$: relaxing the tolerance by $0.01$ would recover about $0.0045$ units of mastery.
The constraint becomes slack near $\delta=0.825$, where $\hat\lambda$ reaches zero.

\paragraph{Comparing criteria.} Table~\ref{tab:case} compares fair policies under several criteria
from Table~\ref{tab:criteria}. Three findings stand out. First, the gap constraint at $\delta=0.6$
is met partly by \emph{levelling down}: the fair policy lowers $\hat V_B$ from $0.482$ to
$0.355$, below the status-quo value $\hat s_B=0.390$, while raising $\hat V_A$ from $-0.337$ to
$-0.245$. Adding a no-harm constraint for both groups makes the problem infeasible at
$\delta=0.6$: no policy in the class closes the gap that far without making group $B$ worse off than
under the status quo. At $\delta=0.7$ both constraints can be met. Second, a floor that guarantees
group $A$ an improvement of $0.15$ over the status quo binds with $\hat V_A=-0.259$, and its
standard error ($0.028$) equals that of $\hat s_A$, as Corollary~\ref{cor:other}(c) predicts:
at a binding floor, $V_A(\thstar)-s_A$ is fixed. Third, the max--min policy maximises
$\hat V_A=-0.235$ but lowers group $B$'s value to $\hat V_B=0.257$, well below the status quo. The standard error of
$\hat V_B$ at this policy is $0.069$ by Corollary~\ref{cor:other}, more than twice the naive
value $0.033$ that treats the policy as fixed. Across the criteria, the naive standard errors
are sometimes larger and sometimes smaller than the correct ones (Table~\ref{S-tab:case-se}), so the
direction of the error is not predictable. Rate parity binds at $|\hat D_A-\hat D_B|=0.10$ but
leaves group $A$'s value essentially unchanged ($\hat V_A=-0.338$ against $-0.337$ without
constraints), which illustrates that parity of decisions and parity of outcomes are different goals.

\paragraph{Certification.} The plug-in fair policy at $\delta=0.6$ has an estimated gap exactly on
the boundary. Its true gap is $-0.001$ (Monte Carlo SE $0.002$), so it is on the boundary to
within simulation error. Proposition~\ref{prop:cert}(a) predicts a violation of order $n^{-1/2}$ in about
half of all samples; in this sample the realised deviation happens to be negligible. The certified policy of
Proposition~\ref{prop:cert}(b) with $\alpha=0.05$ has estimated slack $0.079$ (SE $0.048$),
so its one-sided 95\% lower confidence bound on $V_A-V_B+\delta$ is zero. Its true gap is
$0.068$ (SE $0.002$), and it costs about $0.042$ in estimated value, which is close to the approximation
$\hat\lambda z_{0.95}\widehat{\mathrm{SE}}=0.036$ from Proposition~\ref{prop:cert}(b).

\input{tables/tex_case_table.tex}

%% file: tables/tex_case_table.tex
\begin{table}[t]
\centering
\caption{Synthetic case study ($n=5000$): fair policies under different criteria. Estimates with standard
errors in parentheses: $\hat V$, $\hat V_A$, $\hat V_B$ at the estimated optimum use the influence
functions of Theorem~\ref{thm:eif} and Corollary~\ref{cor:other}; $\widehat\PoF=\hat V(\hat\theta_{\rm unc})-\hat V(\hat\theta)$;
$\hat\lambda$ lists the positive multipliers; $\hat D_A-\hat D_B$ is the difference in the rate of
``advance'' decisions. The last column gives the true value of the estimated policy, which is
known only because the data are synthetic (computed from $2\times10^6$ simulated students; Monte Carlo
standard error about 0.001). Status quo (logging policy): $\hat s_A=-0.409$ (0.028), $\hat s_B=0.390$ (0.027).}
\label{tab:case}
\scriptsize
\setlength{\tabcolsep}{3pt}
\resizebox{\textwidth}{!}{%
\begin{tabular}{@{}lcccccccc@{}}
\toprule
Criterion & $\hat\theta$ & $\hat V$ & $\widehat\PoF$ & $\hat\lambda$ & $\hat V_A$ & $\hat V_B$ & $\hat D_A-\hat D_B$ & True $V(\hat\theta)$\\
\midrule
Unconstrained & (-0.01, -0.09) & 0.194 (0.022) & 0 & -- & -0.337 (0.034) & 0.482 (0.031) & -0.15 & 0.207\\
One-sided gap, $\delta=0.6$\textsuperscript{a} & (0.57, 0.68) & 0.143 (0.028) & 0.050 (0.023) & 0.45 & -0.245 (0.027) & 0.355 (0.027) & -0.12 & 0.181\\
Gap $\delta=0.6$ + no harm & \multicolumn{8}{l}{infeasible: no policy in the class meets both constraints}\\
Gap $\delta=0.7$ + no harm & (0.33, 0.31) & 0.179 (0.022) & 0.015 (0.013) & 0.25 & -0.275 (0.022) & 0.425 (0.022) & -0.13 & 0.204\\
Floor $V_A\ge s_A+0.15$ & (0.44, 0.47) & 0.165 (0.032) & 0.029 (0.024) & 1.11 & -0.259 (0.028) & 0.396 (0.048) & -0.13 & 0.196\\
Max--min & (0.80, 1.15) & 0.084 (0.041) & 0.110 & 1.00 & -0.235 (0.038) & 0.257 (0.069) & -0.09 & 0.138\\
Rate parity, $|D_A-D_B|\le0.1$ & (0.98, -0.51) & 0.158 (0.030) & 0.036 (0.024) & 1.41 & -0.338 (0.053) & 0.428 (0.040) & -0.10 & 0.140\\
\bottomrule
\end{tabular}}

\vspace{2pt}
{\scriptsize \textsuperscript{a}The absolute-gap criterion $|V_A-V_B|\le0.6$ gives the same solution, because its
second constraint is slack.}
\end{table}

%% file: tables/tex_exact_supp.tex
\paragraph{Law.} All variables are binary. $\Pr(G=A)=0.4$; $\Pr(X_1=1\mid G=B)=0.6$ and
$\Pr(X_1=1\mid G=A)=0.35$; $\mu_1(1\mid H_1)=\expit(0.2+0.8X_1-0.5Z)$ with $Z=\1\{G=A\}$;
$\Pr(X_2=1\mid H_1,A_1)=\expit\{-0.4+1.2X_1+A_1(1.4X_1-1.0-0.5Z)\}$;
$\mu_2(1\mid H_2)=\expit(-0.1+0.7X_2-0.4Z+0.3A_1)$; and
$\Pr(R=1\mid H_2,A_2)=\expit\{-0.6+0.9X_2+0.4X_1-0.3Z+A_2(0.8(X_1+X_2)-0.6-0.6Z)+0.3A_1(2X_1-1)\}$.
The policy class is \eqref{eq:softthreshold} with $\kappa=2$, $S_1=X_1$, $S_2=(X_1+X_2)/2$ and
$\theta=(\theta_1,\theta_2)$. The joint law has 64 cells and every functional is a finite sum.

\paragraph{Solution.} The unconstrained optimum is $\theta_{\rm unc}=(0.765,0.524)$ with value
$0.5298$. With $\delta=0.162$ the gap constraint binds: $\thstar=(0.982,1.156)$,
$\lamstar=0.3277$ and $\Vstar=0.5230$. The tangential curvature of the Lagrangian at $\thstar$ is
negative, so Condition~\ref{cond:regular} holds.

\paragraph{Procedure.} For each of 20 directions we drew $s$ uniformly on $[-1,1]$ for each cell,
centred it under $P_0$ and scaled it to $\|s\|_\infty=1$. For $\varepsilon=\pm10^{-5}$ we formed
$P_\varepsilon=P_0(1+\varepsilon s)$, recomputed all conditional laws, and solved the constrained
problem by Newton's method on the KKT system, starting at $(\thstar,\lamstar)$, with gradients
computed by complex-step differentiation (exact to rounding error) and Hessians by central
differences of these gradients; Newton was iterated until the step was below $10^{-15}$. The
unconstrained problem was solved in the same way. The numerical derivative
$\{\Psi(P_\varepsilon)-\Psi(P_{-\varepsilon})\}/(2\varepsilon)$ was compared with $\E_0(\mathrm{IF}\cdot s)$
computed exactly from the 64 cell probabilities, for $\Psi\in\{\Vstar,\thstar,\lamstar,V_A(\thstar),\PoF\}$.
The influence functions of $\thstar$ and $\lamstar$ use \eqref{eq:if-theta-lambda}, with
$\nabla_\theta\bPhi_\theta$ obtained by complex-step differentiation. Table~\ref{tab:exact} in the
main text reports the maximum absolute discrepancies.

%% file: tables/tex_exact_general.tex
\begin{table}[h]
\centering
\caption{General exact checks on a three-group, two-stage discrete law with three policy parameters
and two outcomes (value and intervention rate). Entries are the maximum over 20 random bounded
directions of $|\text{numerical derivative}-\E_0(\mathrm{IF}\,s)|$ (Richardson-extrapolated central
differences), followed by the maximum absolute derivative, for $\Vstar$, the components of $\thstar$,
the active multipliers and the group values $V_g(\thstar)$.}
\label{S-tab:exact-general}
\footnotesize
\resizebox{\textwidth}{!}{%
\begin{tabular}{@{}p{5.2cm}cccccccc@{}}
\toprule
 & \multicolumn{2}{c}{$\Vstar$} & \multicolumn{2}{c}{$\thstar$} & \multicolumn{2}{c}{$\lamstar$} & \multicolumn{2}{c}{$V_g(\thstar)$}\\
\cmidrule(lr){2-3}\cmidrule(lr){4-5}\cmidrule(lr){6-7}\cmidrule(lr){8-9}
Criterion & error & $|$deriv$|$ & error & $|$deriv$|$ & error & $|$deriv$|$ & error & $|$deriv$|$\\
\midrule
I: inequality-averse objective, binding ratio and rate-parity constraints, slack budget & $2.3\times10^{-11}$ & 0.21 & $1.1\times10^{-9}$ & 11.09 & $1.7\times10^{-8}$ & 151.71 & $3.9\times10^{-11}$ & 0.37\\
II: max--min of improvements over the status quo (epigraph), two active constraints & $1.9\times10^{-11}$ & 0.07 & $3.8\times10^{-10}$ & 1.47 & $1.7\times10^{-9}$ & 7.04 & $2.8\times10^{-11}$ & 0.20\\
II in the model with $\mu$ known (restricted scores, known-$\mu$ EIF of $s_g$) & $1.6\times10^{-11}$ & 0.05 & $3.3\times10^{-10}$ & 1.19 & $1.6\times10^{-9}$ & 6.07 & $2.2\times10^{-11}$ & 0.16\\
\bottomrule
\end{tabular}}
\end{table}

%% file: tables/tex_sim_supp.tex
\subsection{Design and implementation details}
The data-generating process is described in Section~\ref{sec:mcdesign}. In a sample of $10^5$ individuals the logged propensities lie in $[0.02,0.97]$; the largest
cumulative weight $\omega_3(\theta)$ is about 6 at the unconstrained optimum and about 11 at
$\thstar(0.5)$, but reaches about 130 at the corners of $\Theta$, which the grid search visits. True values were
computed with $10^6$ simulated individuals and all $2^3$ counterfactual action paths (common random
numbers), so that $V_g(\theta)$ and its gradient are smooth deterministic functions of $\theta$; the
constrained problems were solved by SLSQP and polished by Newton's method on the KKT system.
Table~\ref{S-tab:truth} reports the solutions. In every case the gradient of the
binding constraint is nonzero, the multiplier is positive for $\delta\le0.8$, and the curvature of the
Lagrangian along the constraint surface (last column) is negative, so Condition~\ref{cond:regular}
holds; the Hessian of the Lagrangian is in fact negative definite. Its smallest absolute eigenvalue is
small in the direction of $\theta_1$, so the value surface is flat in that direction and $\hat\theta$
is variable even though $\hV^\star$ is not. The true constraint values and values of estimated
policies (the feasibility and value-loss columns of Tables~\ref{tab:inf}, \ref{S-tab:inf-1000} and
\ref{S-tab:inf-2000}) are evaluated on the same sample of $10^6$ individuals that defines the truth, by bicubic
interpolation of a $51\times51$ grid over $\Theta$;
the maximum interpolation error is below $10^{-7}$. The Monte Carlo standard error
of the truth itself is 0.0013 for values and 0.0025 for the gap $V_A-V_B$.
The tightened problem of Proposition~\ref{prop:cert}(b) had no solution in $\Theta$ (so that no policy
could be certified) in the following numbers of replications: at $n=1000$, 146, 37, 2, 0 replications for $\delta=0.5,\ldots,0.8$, with feasibility 88.4--95.4\% among the others; at $n=2000$, 47, 4, 0, 0 replications for $\delta=0.5,\ldots,0.8$, with feasibility 93.6--97.4\% among the others; at $n=4000$, 6, 0, 0, 0 replications for $\delta=0.5,\ldots,0.8$, with feasibility 92.9--96.4\% among the others.

The outcome regressions use, at stage $t$, the features $(1,G,X_t,X_t^2,GX_t)$ and, for $t\ge2$, also
$(X_{t-1},X_tX_{t-1},A_{t-1})$, each interacted with $(1,A_t)$, fitted by ridge-stabilised least squares
(penalty $10^{-6}n$). In scenarios S2--S4 the propensities are estimated by cross-fitted logistic
regression on $(1,X_t,G,A_{t-1})$ (S2, S3) or on $(1,G)$ only (S4, misspecified); in S3 the outcome model
contains only $(1,A_t)$. All runs used two CPU cores; one replication at $n=4000$ takes about 7 seconds,
including the IPW analysis, the certified policy and the KKT-based influence functions.

\begin{table}[h]
\centering
\caption{True values in the simulation design.}
\label{S-tab:truth}
\footnotesize
\begin{tabular}{@{}cccccccc@{}}
\toprule
$\delta$ & $\thstar$ & $\lamstar$ & $\Vstar(\delta)$ & $V_A(\thstar)$ & $V_B(\thstar)$ & $\PoF(\delta)$ & curvature\\
\midrule
$\infty$ (unc.) & (-0.206, 0.053) & 0 & 0.1814 & -0.3146 & 0.5117 & 0 & --\\
0.5 & (0.954, 0.931) & 0.4122 & 0.1129 & -0.1873 & 0.3127 & 0.0685 & -0.0391\\
0.6 & (0.562, 0.644) & 0.2910 & 0.1481 & -0.2121 & 0.3879 & 0.0333 & -0.0516\\
0.7 & (0.210, 0.379) & 0.1650 & 0.1709 & -0.2493 & 0.4507 & 0.0105 & -0.0588\\
0.8 & (-0.121, 0.121) & 0.0348 & 0.1809 & -0.2993 & 0.5007 & 0.0005 & -0.0597\\
1.0 & (-0.206, 0.053) & 0.0000 & 0.1814 & -0.3146 & 0.5117 & 0.0000 & -0.0589\\
\bottomrule
\end{tabular}
\end{table}

\begin{table}[h]
\centering
\caption{As Table~\ref{tab:inf}, for $n=1000$.}
\label{S-tab:inf-1000}
\footnotesize
\setlength{\tabcolsep}{3pt}
\begin{tabular}{@{}c cc c cccc c cc c cc cc@{}}
\toprule
 & \multicolumn{2}{c}{Cov.\ of $\Vstar$ interval} & & \multicolumn{4}{c}{Multiplier $\lamstar$} & & \multicolumn{2}{c}{Cov.\ $V_A(\thstar)$} & & \multicolumn{2}{c}{$\Pr\{C_1(\hat\theta)\ge0\}$} & \multicolumn{2}{c}{Value loss}\\
\cmidrule(lr){2-3}\cmidrule(lr){5-8}\cmidrule(lr){10-11}\cmidrule(lr){13-14}\cmidrule(lr){15-16}
$\delta$ & Thm 1 & naive & & True & Bias & SD/SE & Cov & & Cor.~1 & naive & & plug-in & certified & plug-in & certified\\
\midrule
0.5 & 97.2 & 93.8 &  & 0.412 & -0.017 & 0.82 & 90.6 &  & 97.2 & 98.4 &  & 40.4 & 62.4 & 0.48 & 6.26\\
0.6 & 96.2 & 95.4 &  & 0.291 & -0.015 & 0.86 & 92.0 &  & 96.2 & 98.8 &  & 43.0 & 86.6 & 0.71 & 6.39\\
0.7 & 96.4 & 96.4 &  & 0.165 & -0.004 & 0.94 & 84.8 &  & 96.0 & 98.8 &  & 47.4 & 94.0 & 0.87 & 4.86\\
0.8 & 96.4 & 96.8 &  & 0.035 & 0.036 & 0.97 & 56.4 &  & 94.2 & 95.6 &  & 61.6 & 95.4 & 0.90 & 2.82\\
\bottomrule
\end{tabular}
\end{table}

\begin{table}[h]
\centering
\caption{As Table~\ref{tab:inf}, for $n=2000$.}
\label{S-tab:inf-2000}
\footnotesize
\setlength{\tabcolsep}{3pt}
\begin{tabular}{@{}c cc c cccc c cc c cc cc@{}}
\toprule
 & \multicolumn{2}{c}{Cov.\ of $\Vstar$ interval} & & \multicolumn{4}{c}{Multiplier $\lamstar$} & & \multicolumn{2}{c}{Cov.\ $V_A(\thstar)$} & & \multicolumn{2}{c}{$\Pr\{C_1(\hat\theta)\ge0\}$} & \multicolumn{2}{c}{Value loss}\\
\cmidrule(lr){2-3}\cmidrule(lr){5-8}\cmidrule(lr){10-11}\cmidrule(lr){13-14}\cmidrule(lr){15-16}
$\delta$ & Thm 1 & naive & & True & Bias & SD/SE & Cov & & Cor.~1 & naive & & plug-in & certified & plug-in & certified\\
\midrule
0.5 & 95.8 & 94.2 &  & 0.412 & -0.002 & 0.78 & 94.4 &  & 96.0 & 97.8 &  & 45.4 & 84.7 & 0.48 & 5.78\\
0.6 & 95.8 & 95.6 &  & 0.291 & -0.001 & 0.96 & 94.4 &  & 95.6 & 98.8 &  & 48.6 & 94.6 & 0.47 & 4.74\\
0.7 & 96.6 & 96.8 &  & 0.165 & 0.001 & 1.00 & 92.4 &  & 96.0 & 98.8 &  & 52.2 & 96.0 & 0.46 & 3.09\\
0.8 & 96.6 & 96.6 &  & 0.035 & 0.025 & 1.06 & 61.8 &  & 93.6 & 95.8 &  & 68.4 & 97.4 & 0.47 & 1.57\\
\bottomrule
\end{tabular}
\end{table}

\subsection{Robustness to nuisance estimation}
Table~\ref{S-tab:robust} compares, at $n=2000$, the main scenario (S1) with three scenarios in which the
logging policy is estimated (300, 300, 300 replications each).

\begin{table}[h]
\centering
\caption{Doubly robust estimator under nuisance misspecification, $n=2000$; entries as in
Table~\ref{tab:main}.}
\label{S-tab:robust}
\footnotesize
\begin{tabular}{@{}l c cccc c cccc@{}}
\toprule
 & & \multicolumn{4}{c}{$\Vstar(\delta)$} & & \multicolumn{4}{c}{$\PoF(\delta)$}\\
\cmidrule(lr){3-6}\cmidrule(lr){8-11}
Scenario & $\delta$ & Bias & SD & SE & Cov & & Bias & SD & SE & Cov\\
\midrule
known $\mu$, series $Q$ & 0.5 & -0.1 & 3.9 & 4.1 & 95.8 &  & 0.3 & 3.4 & 3.5 & 91.0\\
 & 0.6 & -0.1 & 3.5 & 3.6 & 95.8 &  & 0.4 & 2.4 & 2.4 & 90.2\\
 & 0.7 & -0.2 & 3.2 & 3.4 & 96.6 &  & 0.4 & 1.4 & 1.3 & 84.0\\
 & 0.8 & -0.1 & 3.2 & 3.4 & 96.6 &  & 0.3 & 0.6 & 0.5 & 62.0\\
 & 1.0 & 0.2 & 3.3 & 3.5 & 97.0 &  & 0.0 & 0.0 & 0.0 & --\\
\addlinespace
logistic $\hat\mu$, series $Q$ & 0.5 & 0.0 & 3.8 & 4.2 & 96.0 &  & 0.5 & 3.5 & 3.6 & 90.7\\
 & 0.6 & 0.0 & 3.3 & 3.6 & 97.3 &  & 0.5 & 2.4 & 2.4 & 88.0\\
 & 0.7 & 0.1 & 3.1 & 3.4 & 97.3 &  & 0.4 & 1.4 & 1.4 & 83.0\\
 & 0.8 & 0.2 & 3.1 & 3.4 & 96.7 &  & 0.3 & 0.6 & 0.5 & 65.7\\
 & 1.0 & 0.5 & 3.3 & 3.5 & 95.3 &  & 0.0 & 0.0 & 0.0 & --\\
\addlinespace
logistic $\hat\mu$, constant $Q$ & 0.5 & 0.3 & 4.1 & 5.1 & 96.7 &  & 0.5 & 4.3 & 5.2 & 94.7\\
 & 0.6 & 0.4 & 4.3 & 4.6 & 97.0 &  & 0.4 & 3.0 & 3.6 & 90.0\\
 & 0.7 & 0.4 & 4.1 & 4.5 & 97.0 &  & 0.4 & 1.9 & 2.1 & 80.0\\
 & 0.8 & 0.4 & 4.1 & 4.6 & 96.0 &  & 0.4 & 1.2 & 1.0 & 56.3\\
 & 1.0 & 0.8 & 4.4 & 4.9 & 96.7 &  & 0.0 & 0.1 & 0.2 & --\\
\addlinespace
misspecified $\hat\mu$, series $Q$ & 0.5 & 0.4 & 4.9 & 5.1 & 93.7 &  & 0.5 & 4.0 & 4.2 & 92.3\\
 & 0.6 & 0.4 & 4.6 & 4.9 & 95.3 &  & 0.6 & 2.9 & 3.0 & 89.7\\
 & 0.7 & 0.4 & 4.3 & 4.8 & 97.3 &  & 0.6 & 1.9 & 1.8 & 80.3\\
 & 0.8 & 0.5 & 4.2 & 4.8 & 98.3 &  & 0.4 & 0.9 & 0.7 & 59.3\\
 & 1.0 & 0.9 & 4.2 & 4.8 & 98.0 &  & 0.0 & 0.1 & 0.0 & --\\
\bottomrule
\end{tabular}
\end{table}

\subsection{Fairness gap at a fixed policy}
For completeness, Table~\ref{S-tab:gapfixed} reports inference on the fairness gap
$V_A(\theta)-V_B(\theta)+\delta$ at the fixed policy $\theta=\thstar(\delta)$, whose true value is zero for
binding constraints. Coverage is close to nominal (94.0--96.6\%). The mean SE exceeds the Monte Carlo
SD by 3--6\% at $n\le2000$, a known finite-sample feature of cross-fitted variance estimators, and is
within 4\% of it at $n=4000$.

\begin{table}[h]
\centering
\caption{Estimated fairness gap at the fixed policy $\thstar(\delta)$ (doubly robust; $\times100$).}
\label{S-tab:gapfixed}
\footnotesize
\begin{tabular}{@{}cccccc@{}}
\toprule
$n$ & $\delta$ & Mean & SD & SE & Cov (\%)\\
\midrule
1000 & 0.5 & 1.10 & 9.61 & 9.97 & 95.8\\
1000 & 0.6 & 0.93 & 9.19 & 9.42 & 95.0\\
1000 & 0.7 & 0.87 & 8.94 & 9.16 & 95.4\\
1000 & 0.8 & 0.84 & 8.95 & 9.18 & 95.6\\
2000 & 0.5 & 0.21 & 6.68 & 7.05 & 95.0\\
2000 & 0.6 & 0.11 & 6.31 & 6.65 & 95.4\\
2000 & 0.7 & 0.13 & 6.08 & 6.45 & 95.6\\
2000 & 0.8 & 0.14 & 6.10 & 6.45 & 96.6\\
4000 & 0.5 & 0.06 & 5.17 & 4.98 & 94.0\\
4000 & 0.6 & 0.12 & 4.83 & 4.70 & 94.4\\
4000 & 0.7 & 0.19 & 4.63 & 4.56 & 94.6\\
4000 & 0.8 & 0.25 & 4.61 & 4.56 & 94.8\\
\bottomrule
\end{tabular}
\end{table}

%% file: tables/tex_case_supp.tex
\paragraph{Generating model (all data synthetic).} For each of $n=5000$ students, group membership
is $Z=\1\{G=A\}\sim\mathrm{Bernoulli}(0.35)$; engagement $E\sim\mathrm{Bernoulli}(0.6-0.15Z)$; and
initial mastery $X_1\sim N(-0.4Z,1)$. At sessions $t=1,\ldots,4$ the platform chooses $A_t=1$
(advance) with logged probability $\mu_t(1\mid H_t)=0.1+0.8\,\expit(1.5X_t-0.3+0.4E)$, which lies in
$[0.1,0.9]$. The session gain is $Y_t=0.15X_t+A_t(0.5X_t+0.15X_{t-1}+0.2E-0.5Z)+\epsilon_t$, mastery
evolves as $X_{t+1}=0.6X_t+0.2X_{t-1}+0.5Y_t+\nu_{t+1}$ with $\epsilon_t,\nu_t\sim N(0,0.5^2)$ and
$X_0=A_0=0$, and the outcome is $R=X_5$. The policy class is \eqref{eq:softthreshold} with
$\kappa=2$, $S_1=X_1$, $S_t=(X_t+X_{t-1})/2$ for $t\ge2$, a threshold $\theta_1$ for sessions 1--2 and
$\theta_2$ for sessions 3--4, and $\Theta=[-2.5,2.5]^2$. The auxiliary outcome for rate parity is
$R^{(1)}=T^{-1}\sum_tA_t$.

\paragraph{Estimation.} Doubly robust estimates use the logged propensities, $K_{\rm f}=5$ folds and
series regressions on $(1,G,E,X_t,X_t^2,GX_t,EX_t)$ and, for $t\ge2$, $(X_{t-1},X_tX_{t-1},A_{t-1})$,
each interacted with $(1,A_t)$. The same folds are used for the value and rate functionals. Status-quo
values are group means of $R$ with the nonparametric influence function
(Remark~\ref{rem:knownmu}). Constrained problems were solved by SLSQP from the unconstrained
solution and the three best feasible points of a $21\times21$ grid; the max--min criterion uses the
epigraph formulation of Table~\ref{tab:criteria}(h), and the certified policy uses the tightened
constraint of Proposition~\ref{prop:cert}(b) with $\alpha=0.05$. True values were computed by
enumerating the $2^4$ counterfactual action paths for $3\times10^5$ simulated students
for the true price-of-fairness curve; the true values of the estimated policies in
Table~\ref{tab:case} use $2\times10^6$ students in ten independent chunks,
with Monte Carlo standard errors of about 0.001 for values and 0.002 for gaps. The true unconstrained
optimal value is 0.213 and the true
$\PoF(0.6)$ is 0.032.

\input{tables/tex_case_se.tex}

%% file: tables/tex_case_se.tex
\begin{table}[h]
\centering
\caption{Case study: standard errors of functionals at the estimated optimum, computed with the
influence functions of Corollary~\ref{cor:other} (``Cor.'') and naively, treating the estimated
policy as fixed (``naive'').}
\label{S-tab:case-se}
\footnotesize
\resizebox{\textwidth}{!}{%
\begin{tabular}{@{}lcccccccccc@{}}
\toprule
 & \multicolumn{2}{c}{$V_A(\thstar)$} & \multicolumn{2}{c}{$V_B(\thstar)$} & \multicolumn{2}{c}{$D_A(\thstar)$} & \multicolumn{2}{c}{$D_B(\thstar)$} & \multicolumn{2}{c}{$V(\thstar)$}\\
\cmidrule(lr){2-3}\cmidrule(lr){4-5}\cmidrule(lr){6-7}\cmidrule(lr){8-9}\cmidrule(lr){10-11}
Criterion & Cor. & naive & Cor. & naive & Cor. & naive & Cor. & naive & Cor. & naive\\
\midrule
One-sided gap, $\delta=0.6$ & 0.0274 & 0.0335 & 0.0274 & 0.0296 & 0.0339 & 0.0083 & 0.0378 & 0.0070 & 0.0277 & 0.0229\\
Gap $\delta=0.7$ + no harm & 0.0217 & 0.0308 & 0.0217 & 0.0285 & 0.0366 & 0.0077 & 0.0366 & 0.0064 & 0.0223 & 0.0219\\
Floor $V_A\ge s_A+0.15$ & 0.0285 & 0.0318 & 0.0479 & 0.0288 & 0.0419 & 0.0079 & 0.0497 & 0.0066 & 0.0323 & 0.0222\\
Max--min & 0.0379 & 0.0379 & 0.0687 & 0.0332 & 0.0379 & 0.0100 & 0.0570 & 0.0090 & 0.0412 & 0.0255\\
Rate parity & 0.0528 & 0.0376 & 0.0405 & 0.0321 & 0.0563 & 0.0096 & 0.0563 & 0.0075 & 0.0300 & 0.0252\\
\bottomrule
\end{tabular}}
\end{table}